\documentclass[10pt,journal,compsoc]{IEEEtran}
\ifCLASSOPTIONcompsoc
  \usepackage[nocompress]{cite}
\else
  \usepackage{cite}
\fi
\ifCLASSINFOpdf
\else
\fi
\usepackage{graphicx}
\usepackage{amsmath}
\usepackage{xcolor}
\usepackage{soul}

\begin{document}

\newcommand{\sysName}{DrawingInStyles}
%SketchingInStyles
\newcommand{\moduleName}{\emph{SC-StyleGAN}}

\newcommand{\hb}[1]{{\color{cyan}#1}}
\newcommand{\hbc}[1]{{\color{red} [Hongbo: #1]}}
\newcommand{\wanchao}[1]{{\color{black}#1}}
\newcommand{\wanchaos}[1]{{\color{blue} [Wanchao: #1]}}
\newcommand{\wanch}[1]{{\color{red}#1}}
\newcommand{\yh}[1]{{\color{green}#1}}
\newcommand{\yhc}[1]{{\color{pink} [yhc: #1]}}

\title{\sysName: Portrait Image Generation and Editing with Spatially Conditioned StyleGAN}

\author{Wanchao Su, Hui Ye, Shu-Yu Chen, Lin Gao, 
        and Hongbo Fu\thanks{\IEEEauthorrefmark{1} Corresponding author}\IEEEauthorrefmark{1}% <-this % stops a space
\IEEEcompsocitemizethanks{
\IEEEcompsocthanksitem Wanchao Su, Hui Ye, and Hongbo Fu are with the School of Creative Media, City University of Hong Kong.
\protect\\
E-mail: \{wanchao.su, hui.ye, hongbofu\}@cityu.edu.hk
\IEEEcompsocthanksitem Shu-Yu Chen and Lin Gao are with the Beijing Key Laboratory of Mobile Computing and Pervasive Device, Institute of Computing Technology, Chinese Academy of Sciences. Lin Gao is also with University of Chinese Academy of Sciences.
\protect\\
E-mail: \{chenshuyu, gaolin\}@ict.ac.cn
}% <-this % stops an unwanted space
%\thanks{Manuscript received xx xx, xxxx; revised xx xx, xxxx.}
}

\IEEEtitleabstractindextext{%
\begin{abstract}
The research topic of sketch-to-portrait generation has witnessed a boost of progress with deep learning techniques.
The recently proposed StyleGAN architectures achieve state-of-the-art generation ability but the original StyleGAN is not friendly for sketch-based creation due to its unconditional generation nature.  
To address this issue, we propose a direct conditioning strategy to better preserve the spatial information under the StyleGAN framework. 
Specifically, we introduce Spatially Conditioned StyleGAN (\moduleName~for short), which explicitly injects spatial constraints to the original StyleGAN generation process. 
We explore two input modalities, sketches and semantic maps, which together allow users to express desired generation results more precisely and easily.
Based on \moduleName, we present \emph{\sysName}, a novel drawing interface for non-professional users to easily produce high-quality, photo-realistic face images with precise control, either from scratch or editing existing ones. 
Qualitative and quantitative evaluations show the superior generation ability of our method to existing and alternative solutions.
The usability and expressiveness of our system are confirmed by a user study.
\end{abstract}

\begin{IEEEkeywords}
Sketch-based Portrait Generation, Suggestive Interfaces, Data-driven Approaches, StyleGAN, Conditional Generation.
\end{IEEEkeywords}}

\maketitle

\IEEEdisplaynontitleabstractindextext

% \fi
\IEEEpeerreviewmaketitle

\IEEEraisesectionheading{\section{Introduction}\label{sec:introduction}}
% \section{Introduction}
\IEEEPARstart{I}{mage}
generation has been a hot research topic and has drawn much attention in both the computer graphics and the computer vision communities, especially due to the advance of deep learning techniques. 
Remarkable progress emerges for image generation solutions based on deep learning (e.g., generative adversarial networks (GANs)~\cite{Goodfellow2014generative}), in terms of generation resolution~\cite{karras2017progressive}, subject categories~\cite{brock2018large}, training data sparsity~\cite{karras2020training}, etc.
Among various contents in the image generation tasks, the human portrait is a preferably studied subject due to its great need in various applications. 
Creating human portraits from sketches is a widely adopted solution for designers. 
Image-to-image translation frameworks (e.g.,~\cite{isola2017image, wang2018high}) are commonly adopted for converting sketches to images due to impressive generation ability as well as precise controllability over the generated results.

The recent StyleGAN frameworks ~\cite{karras2019style, karras2017progressive} achieve state-of-the-art generation performance for, in particular, portrait images. 
The StyleGAN synthesis network generate images with latent style vectors.
Different spatial resolution ($4^2 - 1024^2$) layers take the style vectors to control different visual attributes: from high-level attributes (e.g., pose, face shape, etc.), smaller-scale facial features (e.g., hairstyle, eyes open/closed), to the coloring scheme and micro-structure.  
Despite the superior performance of StyleGAN, it suffers from a severe drawback when applied to a portrait creation scenario: due to its unsupervised training mechanism, StyleGAN is not suitable for the spatially conditioned generation setting.
Several works (e.g., \cite{richardson2020encoding,tov2021designing}) have attempted to map an input domain 
% \hbc{what?} 
to the StyleGAN latent style space, achieving the indirect control via the latent space. However, encoding the condition to the style space loses the spatial information, and thus cannot guarantee the spatial constraint to be respected %\wanchao{using which the spatial constraint} is not guaranteed to be respected \wanchao{\st{with these approaches}}
after the generation process. 

\begin{figure*}
    \centering
    \includegraphics[width=\linewidth]{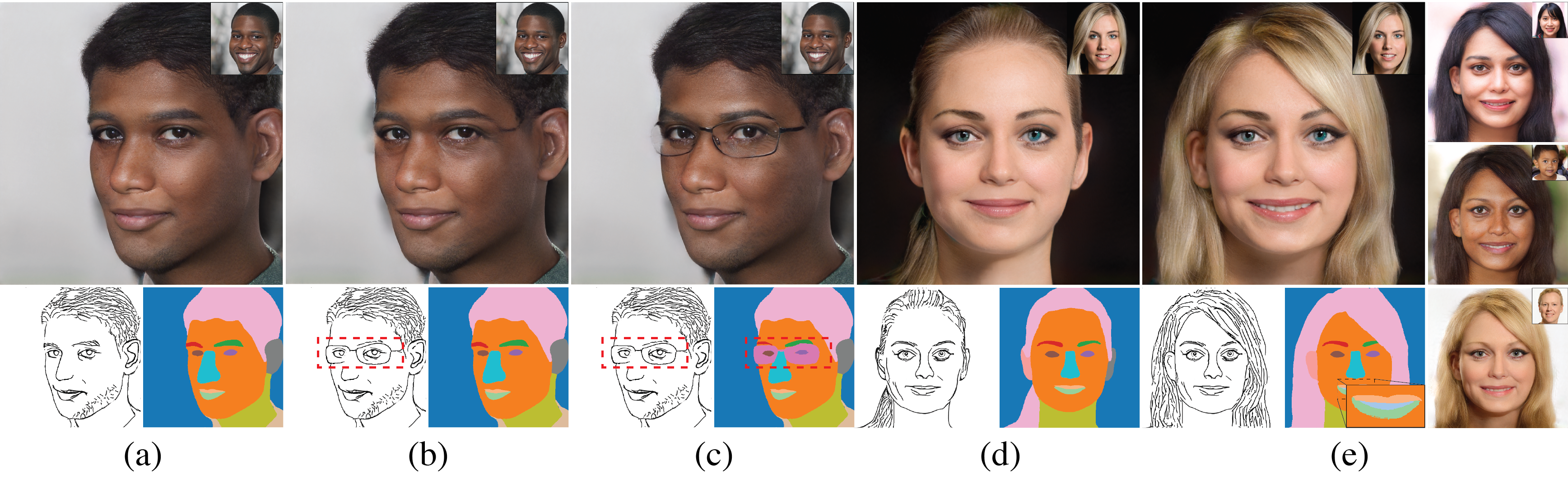} 
    \caption{Our \sysName~ system helps users with limited drawing skills to produce high-quality portrait images with diversified geometry and appearance (best viewed with zoom in) from scratch. Our data-driven suggestive interface assists users in interactive refinement of sketches and semantic maps (Bottom), which provide precise conditions for subsequent image synthesis.
    Our method also supports high-quality portrait image editing (e.g., from (d) to (e): changing the hairstyle, making a smiling face; from (a) to (c): wearing glasses) by editing the sketch and/or semantic map. The minor changes to the input are highlighted in red boxes and zoom-in box. From (b) and (c), it can be seen that the semantic map helps resolve the ambiguity in the sketch, leading to a more expected result.
    }
    \label{fig_teaser}
\end{figure*}

To utilize StyleGAN's ability in portrait image generation, we need to provide a precise control regarding the spatial conditions.
Instead of encoding the spatial conditions into the spatially-oblivious compact style codes, we propose a more aggressive way that transforms the spatial conditions directly into the StyleGAN synthesis procedure.
Since an efficient way to preserve the condition information is to maintain the spatial relationships embedded in the input, we propose to use a spatial encoding scheme to transform the information contained in the condition input.
The original StyleGAN produces results by progressively normalizing randomly initialized spatial feature maps with the guidance of the corresponding style codes. %\st{acquires the spatial information  progressively interpreting the style code inputs and adding to the additional information to the intermediate spatial feature maps.}
We propose to eliminate the gap between the condition-encoded feature and the intermediate spatial feature maps in the StyleGAN synthesis procedure and modify the pre-trained StyleGAN synthesis network as an image-to-image translation architecture.

To achieve the above goal, we present~\moduleName~(Spatially Conditioned StyleGAN
% \hbc{Maybe `Spatially Conditioned StyleGAN'? Update all the instances}
), which consists of a spatial encoding module, a spatial mapping module, and subsequent pre-trained StyleGAN blocks to translate the spatial condition information to high-quality, large-sized ($1024\times1024$) portrait images. 
We encode the input condition to a spatial feature map and then process it with a mapping module before connecting to the subsequent pre-trained StyleGAN synthesizing flow.
We use the encoded feature to substitute the original intermediates guided by the early-stage style codes in StyleGAN. 
This makes the input information a spatial constraint in the generation process.
By training the weights in the encoding and mapping modules, and fixing the pre-trained weights of the subsequent StyleGAN blocks in our \moduleName, we smoothly transform the input condition into the intermediate spatial feature space, thus converting the unconditional StyleGAN synthesis network to a precise and efficient image-to-image synthesis module in our system.

The existing sketch-to-image techniques can be classified into two groups: one requiring accurate sketches as input (e.g., pix2pixHD~\cite{wang2018high}) and one allowing rough/incomplete input (e.g., DeepFaceDrawing~\cite{chenDeepFaceDrawing2020}). The latter is more friendly for novices but lacks precise control (see the comparison between DeepFaceDrawing and ours in Figure \ref{fig_generation}).
Our \emph{\sysName}~falls in the first group and aims to improve the generation quality (see the comparison between pix2pixHD and ours in Figure \ref{fig_generation}).
To fill the gap between these two groups, we propose a suggestive interface, which helps input rough strokes for retrieving edge-map-like global face templates for referencing and face components for explicit refinement. %explicitly refining. 
% The interface assists users in creating  face sketches from scratch by interactively retrieving and presenting face sketches as global templates and face component sketches for refining local details. 

We observe that using sketches and semantic maps together allows users to express themselves more precisely. Based on this key observation and \moduleName, we present \emph{\sysName}, a novel drawing-based system that allows non-professional users to create high-quality face images from sketches and semantic maps, with great ease and precise control (Figure \ref{fig_teaser}). 
Our system can be used for editing portrait images via sketches and/or semantic maps.
Due to the adoption of the StyleGAN architecture, our system supports the change of appearance style for generated results with respect to given reference styles, thus greatly enhancing result diversity (Figure \ref{fig_teaser}).

We compare our system with existing and alternative solutions both quantitatively and qualitatively.
The evaluations prove that our system produces visually more pleasing portrait images. 
The usability of our interface and the expressiveness of our tool are confirmed by a user study. 
We show that our proposed \moduleName conditioning scheme can be further applied beyond the current facial pre-trained model, and demonstrate its extension to the LSUN \emph{Car} and \emph{Church} data~\cite{yu2015lsun}. 

\section{Related Work}

Our work is closely related to the topics of sketch-based portrait generation, portrait image editing with spatial guidance, and StyleGAN manipulation and conditioning.
For each topic, we discuss only the most related works to ours, since a comprehensive review on such topics is beyond the scope of this paper.

\subsection{Sketch-based Portrait Generation}\label{related_SBPG}
Recently, Generative Adversarial Networks (GANs)~\cite{Goodfellow2014generative} and their variations like conditional-GANs \cite{mirza2014conditional} have been widely adopted as generative models for image generation problems. 
For example, \emph{pix2pix} proposed by Isola et al. ~\cite{isola2017image} has become the backbone frameworks for various image-and-image translation problems.
\emph{pix2pixHD} by ~\cite{wang2018high} improves the performance of \emph{pix2pix} and generates higher-resolution results given condition images.
\emph{Scribbler} by Sangkloy et al.~\cite{sangkloy2017scribbler} takes as input sketches and colorizes them under the guidance of user-specified color strokes. 
Such methods and subsequent works (e.g., \cite{zhu2017unpaired, wang2018high}) directly based on them generate results in a pixel-wise correspondence manner, which is similar to our \moduleName. However, our \moduleName~generates results with higher quality (see comparisons in Section \ref{sec_exp_generation}) and supports easy change of coloring and texture details {due to the adoption of StyleGAN framework}. 
Limited by the pixel-wise correspondence nature of the above methods, they require input sketches to be highly similar to the edge maps used for model training to generate quality results, and thus they are not friendly for users with little drawing skills. 
We circumvent this issue by incorporating a data-driven suggestive drawing interface, thus allowing users to quickly find template sketches and update individual face components interactively.

Several attempts have been made towards generating images from imperfect sketches. 
For example, \emph{LinesToFacePhoto} by Li et al.~\cite{li2019linestofacephoto} trains a conditional-GAN embedded with a self-attention module to solve the input incompleteness issue.
Li et al.~\cite{li2020deepfacepencil} employ a spatial attention pooling module to implicitly convert a deformed semantic boundary to the data flow trained with an edge-aligned input, to get a realistic face image. 
Yang et al. ~\cite{yang2020deepps} process a freehand sketch via multi-scale dilation operations, which encode a potential stroke field, and then use a refinement module to get a predicted complete sketch.
Although the above methods have a better ability in handling imperfect sketches than \emph{pix2pix}~\cite{isola2017image}, their ability of handling freehand sketches and generating quality is still limited.
\emph{DeepFaceDrawing} by Chen et al. ~\cite{chenDeepFaceDrawing2020} achieves the state-of-the-art performance for the task of generating realistic face images from rough sketches. 
The key to their solution is the projection of an input rough sketch to component-level manifolds for sketch refinement before the image generation process.
However, the refined sketch is implicitly encoded in this process and users have to control the generated results by interactively updating the rough sketch, instead of the  intermediate features, thus losing precise control of final results.
%These issues are more serious when drawing faces under different poses. This somewhat explains why \emph{DeepFaceDrawing} currently supports the drawing of frontal faces only.
We take a different route from such methods by separating the sketch refinement and image synthesis procedures: we provide a novel interface 
% \hbc{to claim it's novel, better to review the existing drawing assistance tools? If you think this section is already long, you can cover them maybe in the section where you introduce the interface?} 
for users to interactively and explicitly refine an input sketch before sending it to the image generation module. Compared to \emph{DeepFaceDrawing}'s implicit all-in-one learning process for sketch correction and image generation, our method provides more accurate control and presents higher quality of the generated results (see comparisons in Figure \ref{fig_generation}).

\wanchao{In addition, \emph{DeepFaceDrawing} requires a set of aligned faces under the same poses for learning the component-level manifolds and \emph{DeepFaceDrawing} has been demonstrated for generating frontal faces only. The current implementation of \emph{DeepFaceDrawing} handles side-view face generation poorly, as shown in Figure 2 in the supplemental material.  
Extending \emph{DeepFaceDrawing} to handle non-frontal face generation (e.g., by preparing properly sets of training data) is possible but the lack of precise control will still be an issue. 
}

% \yh{A group of the previous works provide suggestive guidance \cite{lee2011shadowdraw, chenDeepFaceDrawing2020} or take a step further to directly correct the imperfections \cite{xie2014portraitsketch} to assist users in sketching faces. Compared to these methods, our proposed interface utilize the global and local retrieval and combine the sketch and semantic map modification.}

\subsection{Portrait Image Editing with Spatial Guidance}
Image editing aims to change certain target regions or attributes of an image according to user inputs while keeping the rest of the image intact and presenting an overall compatible visual appearance.
Here we only focus on the deep learning based image editing works using spatial editing guidance.

Park et al.~\cite{park2019semantic} propose a spatially-adaptive normalization layer for synthesizing photo-realistic images given an input semantic layout. 
Their system supports effective image editing via changing the semantic map of a target image.
Gu et al. ~\cite{gu2019mask}, Lee et al. ~\cite{CelebAMask-HQ}, and Zhu et al. ~\cite{zhu2020sean} propose systems enabling face-component shape editing via altering the semantic map of a target face. 
They also support the control of the target face's appearance by providing encoded features of a reference appearance image with an associated semantic map and applying them back according to the original face map.
Similar to the above methods, our system also enables users to edit the face component shapes according to the edited semantic map.
However, due to the adoption of StyleGAN in our framework, changing the appearance of the target image requires only a reference style code, rather than a face image together with its corresponding semantic map. 
In addition, we propose to use sketches together with semantic maps, since the former is more flexible for specifying local geometric details.

Another widely adopted medium for image editing is the sketch. 
Sketch-based image editing often employs the design idea of sketch-guided image inpainting, which fills a target missing area with the structure provided by an input sketch while referencing the neighboring known areas in generating the textures and colors of the missing area to get the final results.
The method of Yang et al. ~\cite{yang2020deepps} follows this idea for sketch-based face editing but the control preciseness and generation quality are limited since their method is designed for tolerating drawing errors.
FaceShop by Portenier et al.~\cite{portenier2018faceshop} and SC-FEGAN by Jo and Park~\cite{jo2019sc} also adopt the idea of image inpainting and present high-quality face editing results with simple guiding sketches and colored strokes within local regions. A similar idea is adopted by Yu et al. ~\cite{yu2019free} to achieve image completion with mask and sketch guidance.
Although previous works have proposed various mechanisms to improve the compatibility between the synthesized and untouched regions, their results might still exhibit incompatibility artifacts. 
Another problem is that since the textures and coloring details of the region of interest are obtained from the neighboring regions, when the editing area grows, less referencing information remains, which further deteriorates the resulting quality.
Our method takes a different path and achieves face editing by directly modifying the sketch and/or the semantic map derived from an input image, leaving the coloring and texture details synthesized by the subsequent StyleGAN layers using the reference style codes derived from the original image, thus ensuring the global compatibility as well as the detail faithfulness.
% \hbc{derived from the original image?}\wanchaos{currently we have not implement style code inversion procedure, thus I didn't explicitlt address the style of the editing source. I plan to address it as a future work in discussion.}\hbc{You mean it might be difficult to preserve the appearance of the original image after editing?}\hbc{Anyway, don't simply answer my questions. Try to address them by modifying the text. For trivial comments, you may highlight your changes and remove my original comments.}

%Recent 
\emph{DeepFaceEditing} proposed by Chen et al. ~\cite{chenDeepFaceEditing2021}
% , a very recent work concurrent to ours, 
separates the appearance and geometry in a local-to-global manner and achieves state-of-the-art portrait image editing performance. 
It provides a unified framework to extract the geometric representation from both sketches and real images, and to obtain the appearance from another network.
\emph{DeepFaceEditing} adopts a cycle-consistent manner in training the network for synthesizing results from the disentangled appearance and geometry.
Our \emph{\sysName}~ leverages the novel \moduleName~ to encode the geometric information from sketches {and semantic maps} and utilizes the style codes injected to the subsequent StyleGAN layers in synthesizing the appearance. 
Compared to \emph{DeepFaceEditing}, our method provides higher-quality generation results with more variants (e.g. poses, accessories, etc.), as shown in Figure \ref{fig_editing_comp}. 

Previous methods incorporate multi-modality inputs in either heterogeneous or homogeneous way.
Gu et al. ~\cite{gu2019mask}, Lee et al. ~\cite{CelebAMask-HQ}, Zhu et al. ~\cite{zhu2020sean} and Chen et al. ~\cite{chenDeepFaceEditing2021} used multiple input in a heterogeneous way, they extract geometry and appearance information from different sources.
For Portenier et al.~\cite{portenier2018faceshop}, Jo and Park~\cite{jo2019sc}, Yu et al. ~\cite{yu2019free} and our method adopt a homogeneous way. In these method, inputs of various forms provide complementary information of each other, which lead to a better representation of the target.
We propose to use both sketches and semantic maps based on the observation that semantic maps are efficient in defining regions while sketches are suitable for representing structures.

\begin{figure*}
  \centering
  \includegraphics[width=\linewidth]{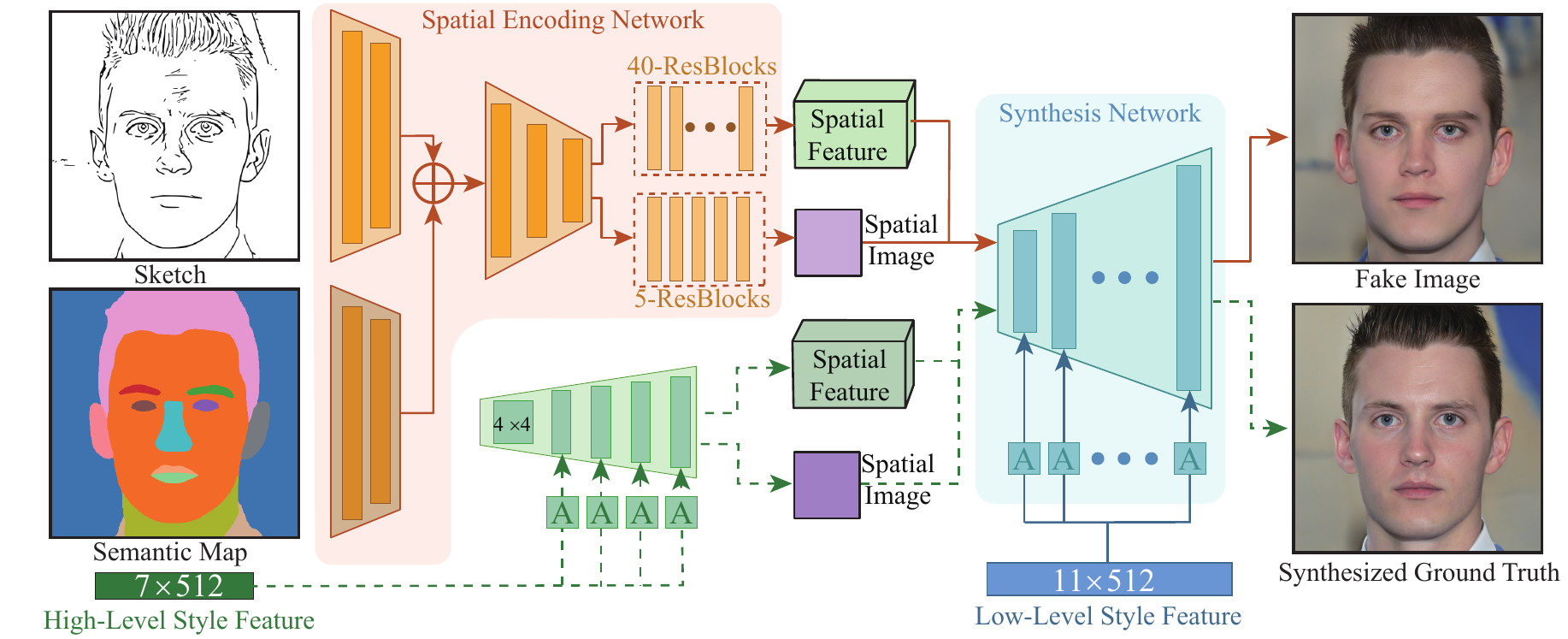}
  \caption{Network architecture of \moduleName. 
  Our \moduleName~consists of a spatial encoding module (in orange) and a subsequent pre-trained StyleGAN synthesis module (in cyan).
  We feed the input sketch and semantic map into individual encoding blocks before being merged using concatenation and fed to a uniform encoding branch to get the spatial feature. 
  The spatial feature is further processed by two separate branches with 40 and 5 ResNet blocks to produce spatial feature map (in light green) and spatial image (in light purple). 
  The spatial encoding module aims to replace the intermediate feature map (in dark green) and the intermediate image (in dark purple) by the original pre-trained high-level synthesis sub-network (in green) with the counterparts encoded from the sketch and the semantic map.
  Our network takes as input a sketch and a semantic map with the paired high-level style features ($7 \times 512$) and a randomly selected low-level style features ($11 \times 512$) from the dataset to obtain the synthesized ground truth in guiding the spatial encoding module to converge.
  The synthesized fake image is produced according the workflow indicated with the solid orange lines and the synthesized ground truth is generated following the dashed green line flow.
  }
  \label{fig_architecture}
\end{figure*}

\subsection{StyleGAN Manipulation and Conditioning}\label{related_GwS}
The portrait image generation quality has been improved over the past years, and the recently proposed StyleGANs~\cite{karras2019style, karras2020analyzing} achieve state-of-the-art visual quality. 
To utilize the rich semantic information in the latent space and exploit the superior generation ability, numerous methods have been developed on top of the StyleGAN architecture and achieve remarkable progress for various semantic manipulation. 
The manipulations are achieved by latent space analysis ~\cite{harkonen2020ganspace, wu2020stylespace}, utilizing pre-trained classifiers ~\cite{shen2020interfacegan, abdal2020styleflow}, controlling a 3D morphable model \cite{tewari2020stylerig}, etc.
Different from the above methods for high-level semantic manipulation, ours applies a more direct spatial control over the generated results, achieving the pixel-wise conditioning for the StyleGAN. 
To achieve semantic manipulation on a given image, the existing methods (e.g. \cite{tewari2020stylerig, shen2020interfacegan}) invert the image to the style latent space, and then apply the designed operations to the inverted latent style codes. 
Existing inversion methods can be roughly divided into three categories: directly optimizing the latent code to minimize the distance to a reference image \cite{abdal2019image2stylegan, abdal2020image2styleganpp}, mapping an image to a latent code \cite{creswell2018inverting, richardson2020encoding}, and the hybrid of the two ~\cite{zhu2020domain}. 
The first category of inversion methods finds style codes in the latent space from a random or projected starting point, thus requiring further time and computation for input inversion.

The second category of the above mentioned inversion methods is a promising route to be modified as an image-to-image translation architecture.
Richardson et al. \cite{richardson2020encoding} propose an image-to-image translation framework called \emph{pixel2style2pixel} (\emph{pSp}), which extends an encoder network from the StyleGAN synthesis module and produces high-quality image embedding results. 
They extract the coarse-to-fine levels of features with a feature pyramid network, and then channel the extracted features to the StyleGAN synthesis layers to obtain translated results. 

Several attempts have been made to support the StyleGAN spatial control. 
Alharbi and Wonka \cite{alharbi2020disentangled} feed multiple noise codes through individual fully-connected layers to spatial noise inputs to control specific parts of generated images. 
StyleMapGAN by Kim et al. \cite{kim2021exploiting} converts the style codes to spatial feature map 
% \hbc{are you sure `stylemap' is a single word?} 
in guiding the normalization process in StyleGAN synthesis.
\wanchao{Barbershop by Zhu et al.\cite{zhu2021barbershop} presents a novel latent space for different sources and fuses the source latent codes according to the semantic mask for image blending. Different from Barbershop,}
our \moduleName~ encodes a sketch image to a spatial feature map before directly injecting it to the spatial layer of the StyleGAN synthesis network, replacing the early stage of the coarse feature map generation process, to transform a sketch image to a portrait image.
The spatial feature map better preserves the stroke information than the compact style code and thus presents better sketch-image corresponding relations (see a comparison with pSp in Figure \ref{fig_generation}).
\section{Methodology}
In this section, we elaborate the details of the portrait image generation process. 
We first introduce the proposed \moduleName~ architecture (Section \ref{mtd_arc}), then present the objective function (Section \ref{mtd_oj}) and finally elaborate the network training strategy (Section \ref{mtd_trn}).

\subsection{\moduleName~Architecture}\label{mtd_arc}
\textbf{StyleGAN}
The original StyleGAN synthesis network takes an $18 \times 512$ style code to its corresponding 18 input layers and generates a high-quality image.
Its synthesis process starts from a randomly initialized constant feature map of spatial resolution $4 \times 4$ and grows by the factor of 2 with the upsampling operations, and finally get a $1024 \times 1024$ resulting image.
In the progressive generation process, each style block takes as input a $1 \times 512$ style code in the transformation of the weights, which are associated with the subsequent convolution operation to control the generation process. 
StyleGAN employs this mechanism to control the generated attributes with the style code inputs.
The original paper of StyleGAN \cite{karras2019style} illustrates the effects for coarse ($4^2-8^2$), middle ($16^2-32^2$), and fine ($64^2-1024^2$) styles, which correspond to the high-level attributes (e.g., pose, face shape, etc.), smaller-scale facial features (e.g., hair style, eyes open/closed), and the coloring scheme and micro-structure, respectively.

\textbf{\moduleName}
To achieve our conditional generation goal, we incorporate the sketch and semantic map to determine the spatial attributes, which suit the purposes of the coarse and middle styles of the original StyleGAN (``High-Level Style Feature'' in Figure \ref{fig_architecture}). 
As illustrated in Figure \ref{fig_architecture}, our \moduleName~consists of two sub-networks: the \emph{spatial encoding} network aims to map the input conditions to intermediates corresponding to the results of the coarse and middle style controlled layers; the \emph{synthesis} network utilizes the pre-trained layers of the original StyleGAN synthesis network and takes as input our spatial encoded intermediates to generate a synthesized image.

Specifically, in our spatial encoding network, we propose two encoding modules that map the $512\times512$ sketch and the $512\times512$ semantic map to spatial feature maps of size $64\times256\times256$ independently.
The resulting two feature maps are concatenated in the channel dimension, resulting a combination (size $128\times256\times256$) of the two modalities of conditions before going through the subsequent encoding process.
The combined feature map is encoded to the spatial resolution of $32\times 32$, which matches the size of the feature map in the StyleGAN synthesis  module in the coarse to middle styles ($4^2-32^2$). 
Before sending to the \emph{synthesis} network, we pass the feature map to 40 ResNet blocks (the upper orange dashed rectangle in Figure \ref{fig_architecture}) to make it better match the intermediate feature map in the original synthesis network.  
Similar to the procedure of producing the spatial intermediate feature map, we propose another branch of 5 ResNet blocks from the embedded feature map (the lower orange dashed rectangle in Figure \ref{fig_architecture}) to generate a $32\times32$ intermediate image, which also matches the counter part in the original StyleGAN synthesis module. 
In the subsequent generation process, we replace the intermediate feature map and image with the feature map and image embedded using our spatial encoding network, as illustrated in Figure \ref{fig_architecture}. 
Each encoding block consists of one convolution layer with stride 2, leaky ReLU activation, and a normalization layer.

\subsection{Objective Function}\label{mtd_oj}

Since the goal of our \moduleName~ is to encode the spatial constraints for the StyleGAN synthesis process while preserving the generation quality of the pre-trained StyleGAN, we need to precisely map the encoded condition to its counter parts in the original synthesis process. 
To achieve this, we formulate the objective function of the training process as follows:
\begin{equation}
    \begin{split}
        L(I_{gt}, I_{syn}) = & \lambda_{L_1}L_1(I_{gt}, I_{syn}) + \lambda_{L_{GP}}L_{GP} \\
         + & \lambda_{L_{LP}}L_{LP} + \lambda_{L_{FM}}L_{FM},
    \end{split}
    \label{eq_oj}
\end{equation}
where $L_1(\cdot, \cdot)$ is the mean abstract difference function, $L_{GP}$ and $L_{LP}$ stand for the global perceptive loss and local perceptive loss, respectively. 
The original StyleGAN uses an adversarial loss in guiding the network convergence, here we use the pre-trained StyleGAN and aim to guide our encoding module converging to the original intermediate space.

\begin{figure}
    \centering
    \includegraphics[width=0.98\linewidth]{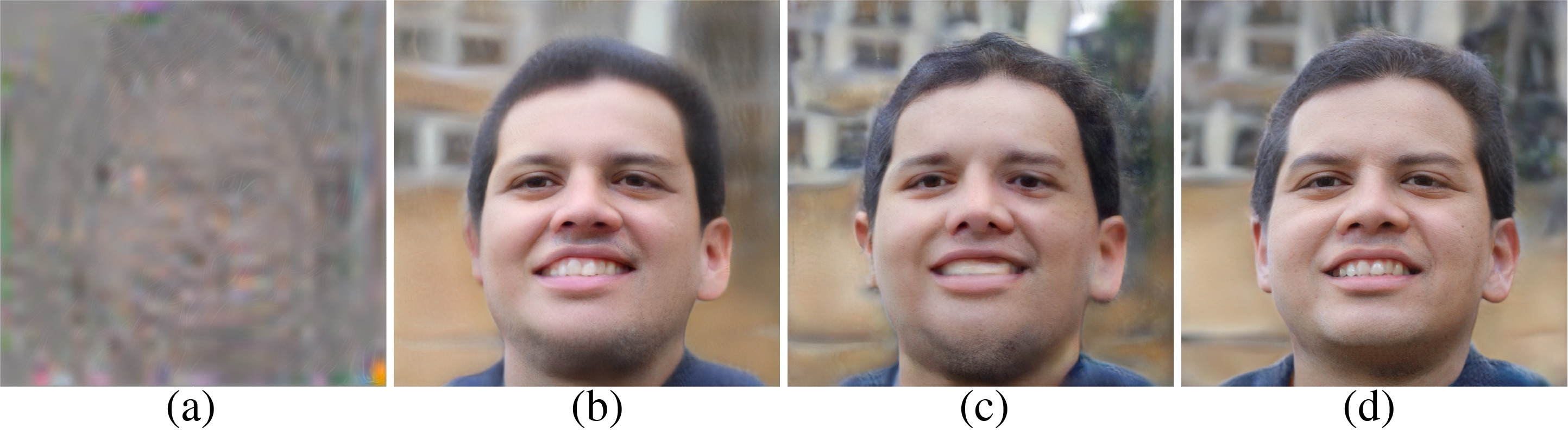}
    \caption{
    Illustration of the effects of the components in the objective function. (a): No $L_1$ Loss, (b): No Perceptual Loss, (c): No Local Perceptual Loss, and (d): Full Method.
    }
    \label{fig_ablation}
\end{figure}

We incorporate the perceptive loss in our objective function to enhance the guidance of the synthesis process.
Since we replace the spatial intermediates in the original StyleGAN workflow with our encoded counterparts, the volume of the optimization targets (i.e., intermediate feature maps and intermediate image) is larger than that of the existing latent space optimization methods. 
Thus apart from the commonly used perceptual loss applied over the full scale of the generated image, we further incorporate a perceptual loss in the local patches.
Inspired by \cite{park2020contrastive}, we randomly crop K patches (K = 20 in training) from the generated and ground-truth images and compute the local perceptual loss.
\wanchao{Here we chose K = 20 to balance the computation cost and final generation quality: when K $>$ 20, our method showed no further quality gain; when K $<$ 20, we experienced quality degeneration.}

We measure the global perceptive loss by resizing the synthesized and target images to spatial size $64\times64$, and measure them with the perceptual metric (LPIPS~\cite{zhang2018perceptual}). Mathematically, the global perceptive loss $L_{GP}$ and the local perceptive loss $L_{LP}$ are formulated as follows:  
\begin{equation}
    \begin{split}
        L_{GP}(I_{gt}, I_{syn}) & = LPIPS(I_{gt}^{re}, I_{syn}^{re}), \\
        L_{LP}(I_{gt}, I_{syn}) & = \frac{1}{K}\sum_{k=1}^{K}LPIPS(I_{gt}^k, I_{syn}^k), 
    \end{split}
    \label{eq_pcpt}
\end{equation}
where $LPIPS(\cdot, \cdot)$ represents the perceptual measuring function, $I_{gt}^{re}$ and $I_{syn}^{re}$ are the resized ground truth and synthesized images, respectively.
$I_{gt}^k$ and $I_{syn}^k$ represent the $k$-th randomly cropped ground truth and synthesized patches in each step, respectively.

To further ensure that our synthesized result approximates to the ground truth, we add another feature matching loss in the objective function:
\begin{equation}
    L_{FM} = \frac{1}{N}\sum_{l}{\|G^l(gt) -  G^l(syn)\|}_1, 
    \label{eq_fm}
\end{equation}
where $G^l(\cdot)$ is the $l$-th resolution block (with the corresponding spatial resolution of $2^l$) output feature map of the pre-trained StyleGAN synthesis network. 
$N$ is the number of calculated blocks.
Here we calculate the $L_1$ norm between the synthesized and ground truth generation process after the replacement resolution block ($l\in\{6, 7, 8, 9\}$ and $N=4$).   
Figure \ref{fig_ablation} illustrates the effects of each component in the objective function. For the details of the ablation study, please refer to Section \ref{sec_exp_ablation}. 

\begin{figure}
    \centering
    \includegraphics[width=0.45\textwidth]{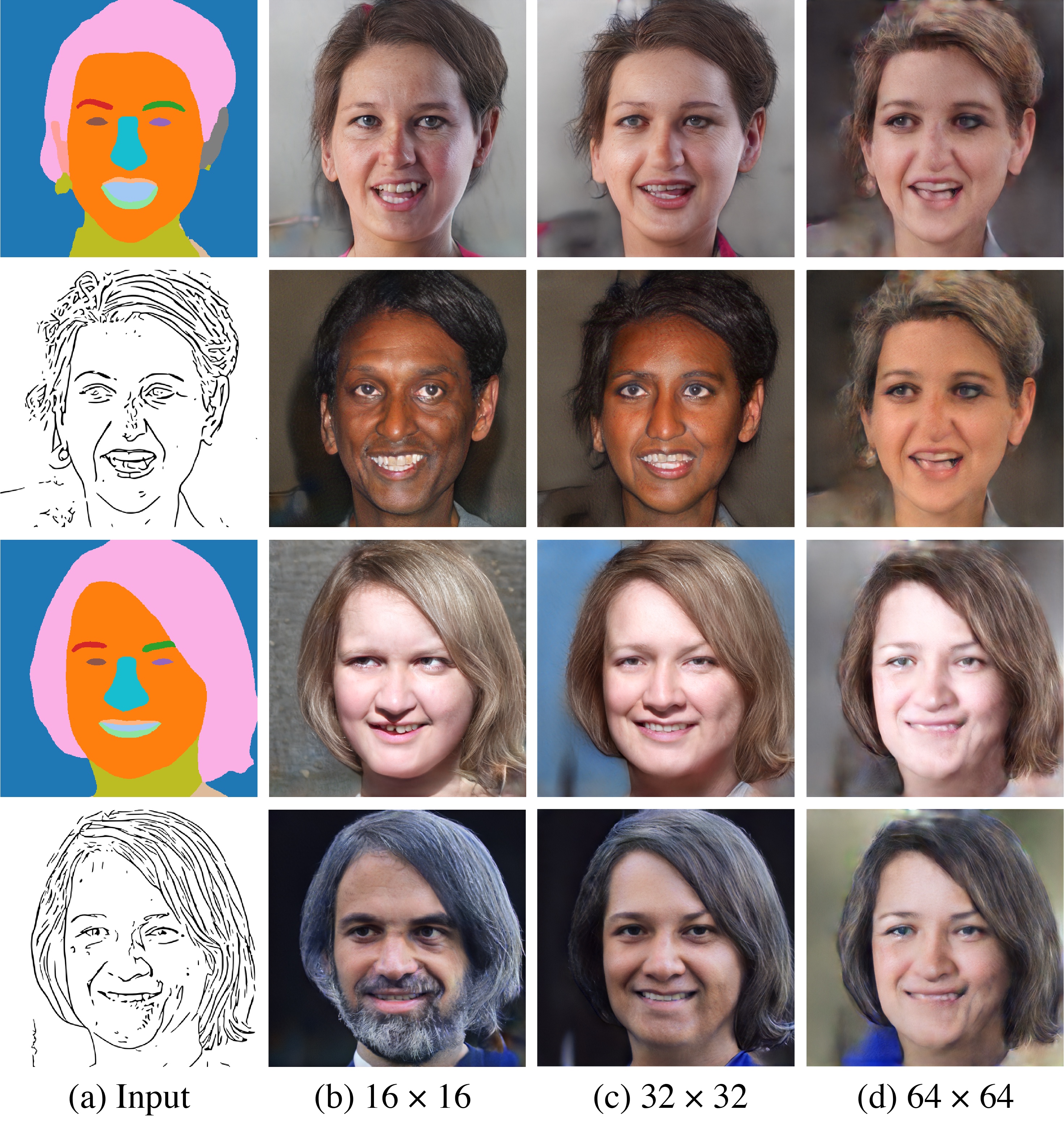}
    \caption{An illustration of different replacement schemes in our \moduleName~ generation process. We show two examples with randomly selected low-level styles.}
    \label{fig_injection}
\end{figure}

\subsection{Training Strategy}\label{mtd_trn}
The training of the \moduleName~is illustrated in Figure \ref{fig_architecture}.
To disentangle the high-level style geometries from the low-level style appearances and augment the existing training dataset, we adopt a dynamic guiding scheme by synthesizing the target image in each iteration.
To achieve this, we generate the synthesized ground-truth target by feeding the pre-trained StyleGAN synthesis network with input-paired high-level style codes ($4^2-32^2$) and randomly selected low-level style codes ($64^2-1024^2$) from the recorded style code dataset.
% (elaborated in Section \ref{sec_data}).
Our \moduleName~synthesizes an image by feeding a sketch and a semantic map paired with the high-level style codes to the \emph{spatial encoding} network to get an intermediate feature map and an intermediate image. 
We then inject the intermediates to the subsequent \emph{synthesis} network with the same low-level style codes ($64^2-1024^2$).
We freeze the parameters of the blocks subsequent to the replacement of the encoded condition intermediates in the \emph{synthesis} network.
We also tried injecting the encoded feature map to different spatial layers (see Figure \ref{fig_injection} for an illustration).
Please refer to Section \ref{sec_exp_ablation} for quantitative evaluation and analysis regarding the different replacement schemes. 

\section{Suggestive Drawing Interface}\label{sec_ui}
To assist users in generating high-quality portrait images with ease and precise control, we propose a data-driven suggestive interface.
Our interface supports image creation from scratch or by editing existing images. The default mode is to create portrait images from scratch. 
It consists of three stages to help non-professional users produce high-quality portrait images, namely, \emph{global selection}, \emph{local detail suggestion}, and \emph{sketch and semantic map modification}. It supports an explicit and coarse-to-fine sketch refinement and mask modification process. To edit an existing image, the user loads an image from the local source, and our system extracts its corresponding sketch and semantic map. In this case, our system automatically skips the global selection and starts from the local detail suggestion stage. Please refer to the accompanying video for the interaction process.

\subsection{System Design}
\textbf{Global Selection}\label{sec_ui_global}
We assist users in globally retrieving relevant faces from the dataset by drawing a coarse contour of a target face. Since novice users are usually not very good at drawing faces with proper proportions, we use the user-drawn strokes in this stage only for retrieval. Our main goal here is to allow users to quickly select a sketch template by simply drawing several strokes.

\begin{figure}
    \centering
    \includegraphics[width=0.9\linewidth]{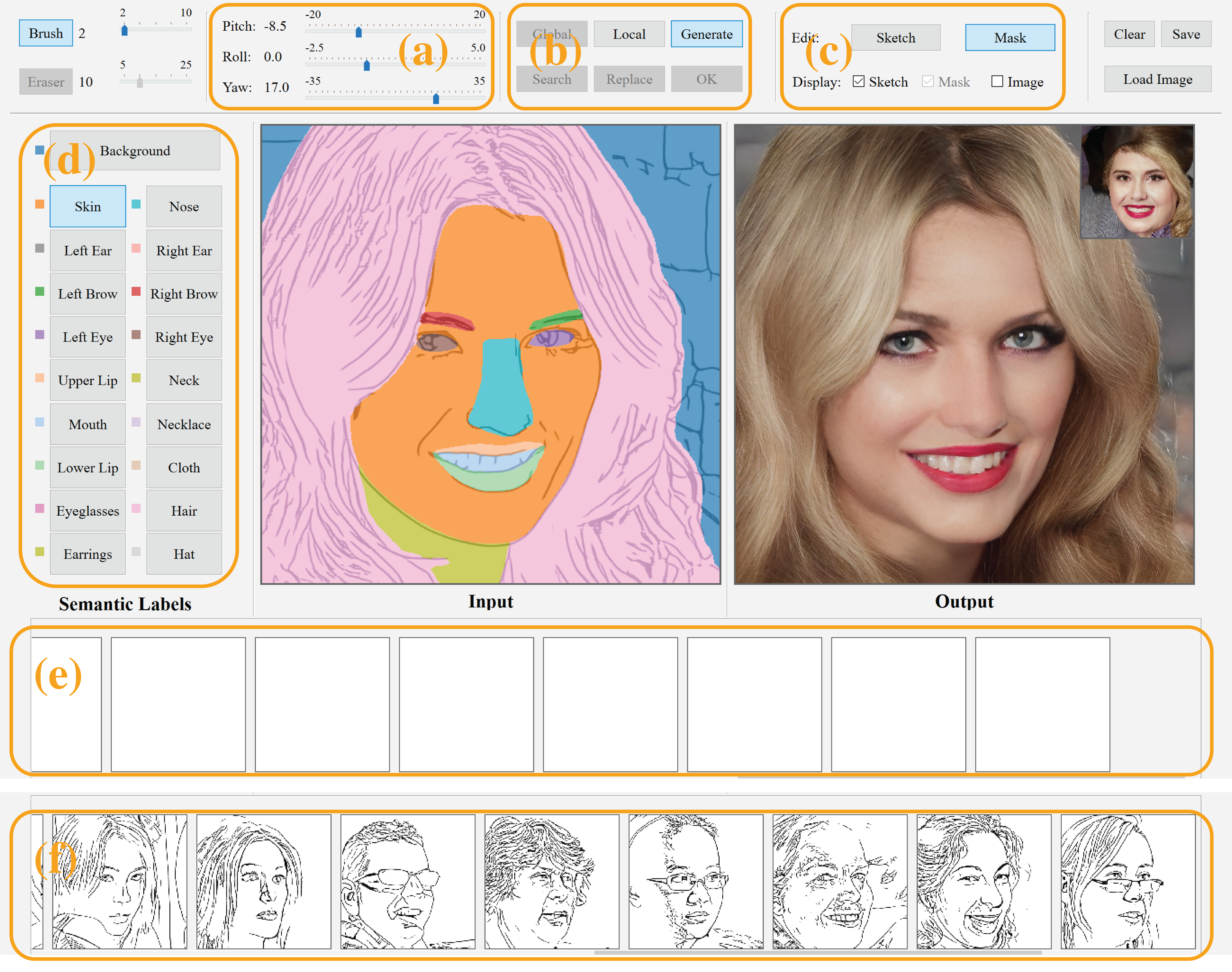}
    \caption{A screenshot of our sketching interface for portrait image synthesis. The sketched strokes and semantic map are displayed in the left canvas. The corresponding synthesized result and a reference style image {(in the top-right corner)} are displayed on the right. 
    The pose sliders on the top (a) are used in the global selection step to find faces under a specific pose. The user can switch the modes of global {selection}, local {detail suggestion}, sketch and mask editing in (b) and (c). The leftmost buttons (d) allow mode switching for selecting, drawing, and editing specific components. The scroll view at the bottom (f) displays the global and local component retrieval candidates, and is empty in the stage of sketch and mask modification (e).} 
    \label{fig_interface}
\end{figure}

Drawing faces under various poses is also challenging for users with little drawing skill. 
To help users easily sketch a face under a specific pose, we provide three pose sliders (Figure \ref{fig_interface})(a) corresponding to Euler angles for 3D rotation to specify a certain pose. Every time a user changes any of the rotation parameters, our system returns a set of face sketches that have their poses as close to the user-specified pose as possible. We extract the contours of the top-20 faces and merge them as one guidance image semi-transparently displayed on the drawing canvas, similar to ShadowDraw \cite{lee2011shadowdraw} and AverageExplorer \cite{zhu2014averageexplorer}. 

When the user draws on top of the guidance, our system re-ranks the face sketches retrieved from the previous step, based on their similarity (Section \ref{retrieval_construction}) to the user-drawn strokes, and displays the top-20 re-ranked face sketches at the bottom of the interface (Figure \ref{fig_interface}(e)).
The user can select one of them and the system shows the selected one to replace the user-drawn strokes in the drawing canvas for further refinement. Each modification will trigger a new re-ranking of the sketch templates. 

\textbf{Local Detail Suggestion}\label{sec_ui_local}
When finishing global selection in the face creation mode or loading an existing image in editing mode, the user can switch to the local detail suggestion stage for component retrieval and modification. In this mode, the user can click on a specific button in the left-most column (Figure \ref{fig_interface}(d)) to select a semantic component label of interest and the system will then display a corresponding red rectangle on the sketched face. 

For each selected component, our system retrieves and displays the top-20 component candidates for selection. For a selected component candidate, it is placed underneath the currently sketched component at the corresponding position in a semi-transparent layer for previewing. The user can either keep the previewed component to replace the current component sketch by clicking on ``Replace'' button or refine the current component sketch according to the previewed guidance. Once finishing the refinement, the user presses the ``OK'' button to remove the component guidance in the sketching canvas. 

During this process, the user can also adjust the position of the individual component sketch/guidance by simply dragging the red rectangle to a desired position. The user can modify the current sketch and start a new retrieval of the component.
The local semantic map will replace the original one with the ``Replace'' button clicked, otherwise, the original semantic map remains intact.

\textbf{Sketch and Semantic Map Modification}
After local component refinement, the user can switch to the sketch and semantic map modification stage. The user can select a specific semantic label (Figure \ref{fig_interface}(d)) to modify the semantic map. In this stage, the user can see the synthesized face image updating in real time in the output canvas. The sketch, semantic map, and result image can be displayed on the sketching canvas for reference by toggling the check-boxes. The user can load a reference image for defining the appearance of the synthesized portrait image. Otherwise a default appearance reference image will be used. 

\subsection{Implementation}\label{retrieval_construction}
We train a ``global contour'' embedding network to construct the global repository for the initial global retrieval.
Similar to DeepFaceDrawing \cite{chenDeepFaceDrawing2020}, we train six component embedding networks, namely, ``facial skin'', ``nose'', ``left-eye", ``right-eye", ``mouth", ``glass'', ``hat'', and ``hair", respectively, to encode the component sketches for constructing the component repositories. 
Each individual embedding network is an auto-encoder architecture. The component sketch goes through the corresponding encoder network via a compact bottleneck layer to get a 512-dimensional compact representation before feeding it to the decoder network.
We adopt a self-supervised learning scheme, which aims to reconstruct the input with a $L_1$ loss as the objective function.
When a query sketch comes, the corresponding trained component encoder first processes it to a compact representation, and then the system uses the resulting representation as query to retrieve the most similar components in the corresponding repository.
Accessories like glasses and hats are directly extracted from the dataset and organized according to their portrait poses. During selection, we simply use the target portrait pose as a query and select portraits with the accessory to obtain the accessory candidates to choose from.

To edit an existing image, we need to obtain its sketch and semantic map.
For the image-to-sketch process, we train a U-net \cite{ronneberger2015u} using our existing sketch-image pairs.
\wanchao{To train our model as a conditional generation framework, we need a large-scale dataset of condition-portrait pairs.
% \hbc{refer to some illustration figure?}. 
A sketch and a semantic map together form the conditions to guide portrait image generation.  
We take advantage of the generation ability of the StyleGAN framework (StyleGAN2 \cite{karras2020analyzing} throughout this paper) in constructing the training dataset by collecting a large series of generation results.
We first sample a large collection of random vectors from a normal distribution before feeding them to the mapping network {of StyleGAN}. 
We then input the resulting latent style codes from the mapping network to the synthesis network and obtain the portrait images corresponding to the style codes.
Up to now, we get a collection of pairs of latent codes and images.}
To get its semantic map, we use BiSeNet \cite{yu2018bisenet} pre-trained on the CelebAMask-HQ dataset \cite{CelebAMask-HQ}. The details of the data preparation can be found in our supplemental materials.
% \wanchaos{Move the data preparation to supplementary.}\hbc{Okay, the paper is already quite long}
Once the user loads an image from an external source, our system sends it to the trained modules and gets the resulting sketch and semantic map for subsequent editing process.
\section{Experiments}
We have conducted extensive experiments to evaluate the effectiveness and usefulness of our method, both quantitatively and qualitatively. 
The experiments were done on a server PC with Intel i7-7700 CPU, 32GB RAM and a single GeForce 1080 Ti GPU. 
Our method generates results with 0.11 second per image on average, and thus supports editing at an interactive rate.
We implemented the suggestive drawing interface and conducted the drawing sessions on a Surface Pro 7 with a Surface Pen. The user inputs and generated results were transmitted between the client and server PC under http protocol. 

In this section, we first show the quantitative results and the analysis of the current architecture and alternative configurations of our method with an ablation study in Section \ref{sec_exp_ablation}.
Comparisons on the generation abilities among different input schemes and alternative methods are introduced in Section \ref{sec_exp_generation}. 
We then compare our method with the state-of-the-art portrait editing solutions in Section \ref{sec_exp_editing}.
The usability of our system is confirmed by a user study, as  elaborated in Section \ref{sec_exp_usability}.
In Section \ref{sec_exp_perceptive}, we further compare the visual quality of results using our method and alternative solutions by a perceptive study. 
We demonstrate that our proposed conditioning ideas can work beyond faces by applying them to the LSUN \emph{Car} and \emph{Church} dataset \cite{yu2015lsun} in Section \ref{sec_exp_other}.
For more generation results, please refer to our accompanying video and the supplemental materials.

\subsection{Ablation Study}\label{sec_exp_ablation}

\begin{table}[t]
    \centering
    \begin{tabular}{ccccccc}
    \hline
         Config & No L1 & No Pcpt & No LP & No GP & No FM & Full \\
    \hline
        L1     &   0.239 &  0.108 &  0.115 &  \textbf{0.095} & 0.100 &  0.098 \\
        Local  &   0.519 &  0.253 &  0.230 &  \textbf{0.177} &  0.185 &  \textbf{0.177} \\
        Global &   0.290 &  0.116 &  0.088 &  0.073 &  0.073 &  \textbf{0.067} \\
        FID    & 378.174 & 56.317 & 40.787 & 33.016 & 35.770 & \textbf{30.265} \\
    \hline
    \end{tabular}
    \caption{\label{tab_ablation} Quantitative results of the ablation study on the terms in the objective function. ``Pcpt'', ``LP'', ``GP'', and ``FM'' mean the perceptual loss in total, local perceptual, global perceptual, and feature matching losses, respectively.
    }
\end{table}

To validate the impact of different terms in our objective function (Equation \ref{eq_oj}), we conducted an ablation study by omitting each component loss in turn in the network training process.
We evaluated the generation results by using the test set as input with the corresponding recorded latent styles, and comparing the reconstructed results with the ground-truth images. We measured the results using $L_1$ loss, local perceptive loss with LPIPS (randomly cropped 20 corresponding patches from both the generated and ground-truth images, as done in the training process), global perceptual loss with LPIPS, and Fr\'echet Inception Distance (FID \cite{heusel2017gans}). Table \ref{tab_ablation} shows the quantitative comparison results.

From Table \ref{tab_ablation} we can see that the $L_1$ loss provides the main optimization direction, as also confirmed by Figure \ref{fig_ablation}(a).
The incorporation of the perceptual losses improves the quality significantly. Without these losses the generated results exhibit blurry artifacts, especially for regions other than the main facial components (e.g., hair region, Figure \ref{fig_ablation}(b)). 
Specifically, the local perceptive loss provides sharp details in the generated results. With the global perceptual loss alone, the generated results may lose fine details (see the blurry mouth region in Figure \ref{fig_ablation}(c)) due to the resizing operation from $1024^2$ to $64^2$. 
The feature matching and global perceptual losses further refine our network optimization. Although the quantitative metrics show limited increments and no significant visual quality improve with these two losses, the incorporation of them accelerates the convergence.

\begin{table}
    \centering
    \begin{tabular}{cccccc}
    \hline
        Config & Mask & Sketch & Both($32\times32$) & $16\times16$ & $64 \times 64$\\
    \hline
        L1     &  0.121 &  0.105 &  \textbf{0.098} &  0.120 &  0.149\\
        Local  &  0.223 &  0.190 &  \textbf{0.177} &  0.217 &  0.261\\
        Global &  0.103 &  0.076 &  \textbf{0.067} &  0.091 &  0.132\\
        FID    & 46.372 & 35.116 & \textbf{30.265} & 33.041 & 62.506\\
    \hline
    \end{tabular}
    \caption{\label{tab_inputs} Quantitative evaluation on the different input choices and replacement schemes.}
\end{table} 

As mentioned in Section \ref{mtd_trn}, we have attempted to replace the intermediates in different spatial resolutions. 
We changed the encoded intermediate sizes by adjusting the number of convolution blocks in the \emph{spatial encoding} network.
We experimented with the replacement in spatial resolutions of $16\times16$ and $32\times32$, and measured the generated results with the same metrics as above. 
We report the quantitative results in Table \ref{tab_inputs} Columns 3--5. 
% Configuration ``Adopted'' represents our adopted replacement in $32\times32$ resolution. 
It can be seen that the current $32\times32$ replacement scheme presents the best performance.
Besides, Figure \ref{fig_injection} illustrates an example of results with the different injection schemes.
 
In this experiment, we fed each individual sample with two sets of randomly selected low-level styles to get the target portrait with different appearances.
We can see that the results of spatial resolution $64 \times 64$ (Figure \ref{fig_injection}(d)) are blurry. This is mainly because larger injection resolution involves more parameters in the encoded intermediates, which is hard to precisely match.
This confirms the longer time consumption in training the model .
This phenomenon also verifies the quantitative results in Table \ref{tab_inputs}.
Despite the reasonable performance of test set reconstruction of the smaller spatial resolution ($16 \times 16$) injection model, the qualitative results present obvious artifacts:
in Figure \ref{fig_injection}(b), we can easily notice that with different low-level styles, the results present high-level semantic changes (e.g., gender change in both cases, beard adding in the second case), which is usually not desirable.
The referencing style codes in the $16 \times 16$ injection setting also contain certain high-level information, and they are interpreted by the subsequent \emph{synthesis} network, thus presenting high-level semantic changes in the example.

\subsection{Evaluation on Generation Performance}\label{sec_exp_generation}
% \begin{figure}
%     \centering
%     \includegraphics[width=0.45\textwidth]{IEEEtran/fig/overlay.png}
%     \caption{
%     Visual comparisons between our method and the state-of-the-art sketch-based image synthesis approaches. We use the low-level styles paired with the sketch to provide the appearance in ours and pSp-ref.
%     }
%     \label{fig_overlay}
% \end{figure}

\begin{figure}
    \centering
    \includegraphics[width=0.44\textwidth]{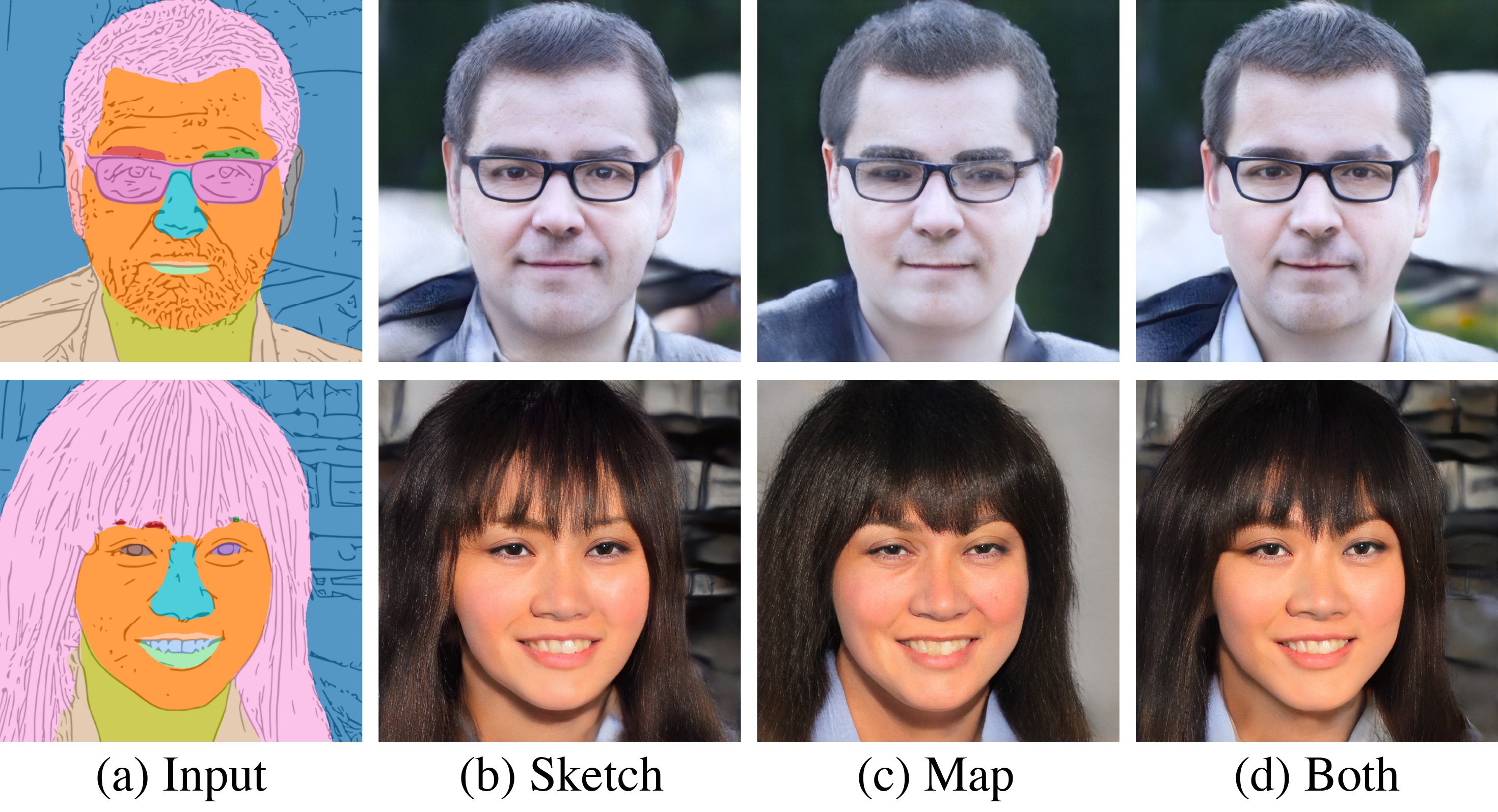}
    \caption{Qualitative comparisons of different input schemes. 
    In the first column, we overlay the two modalities together.}
    \label{fig_inputs}
\end{figure}

We propose to employ both sketch and semantic map in our generation process. In this experiment, we validated our adopted input scheme over the alternatives: sketch only and semantic map only.
We altered the input processing at the beginning of our \emph{spatial encoding} network in \moduleName.
To ensure the dimensionality consistency and comparison fairness, we doubled the output channel dimension of the output module for the case of single input modality to offset the concatenation of two modalities in our adopted configuration.
Qualitative results are shown in Figure \ref{fig_inputs} to demonstrate the difference in results with different inputs and Table \ref{tab_inputs} Columns 1--3 shows the results of quantitative comparisons. 

As we can see, with the two input modalities, our method performs the best in terms of all the metrics, and this verifies the adoption of the two modalities of input benefits the generation.
However, we can notice that with sketch only, our framework also provided reasonable performance.
We attribute this to that the sketch itself can provide both region boundary and structure information, while incorporating the semantic map further eliminates the ambiguity when the network interpreting the input sketches.

\begin{figure*}
    \centering
    \includegraphics[width=0.99\linewidth]{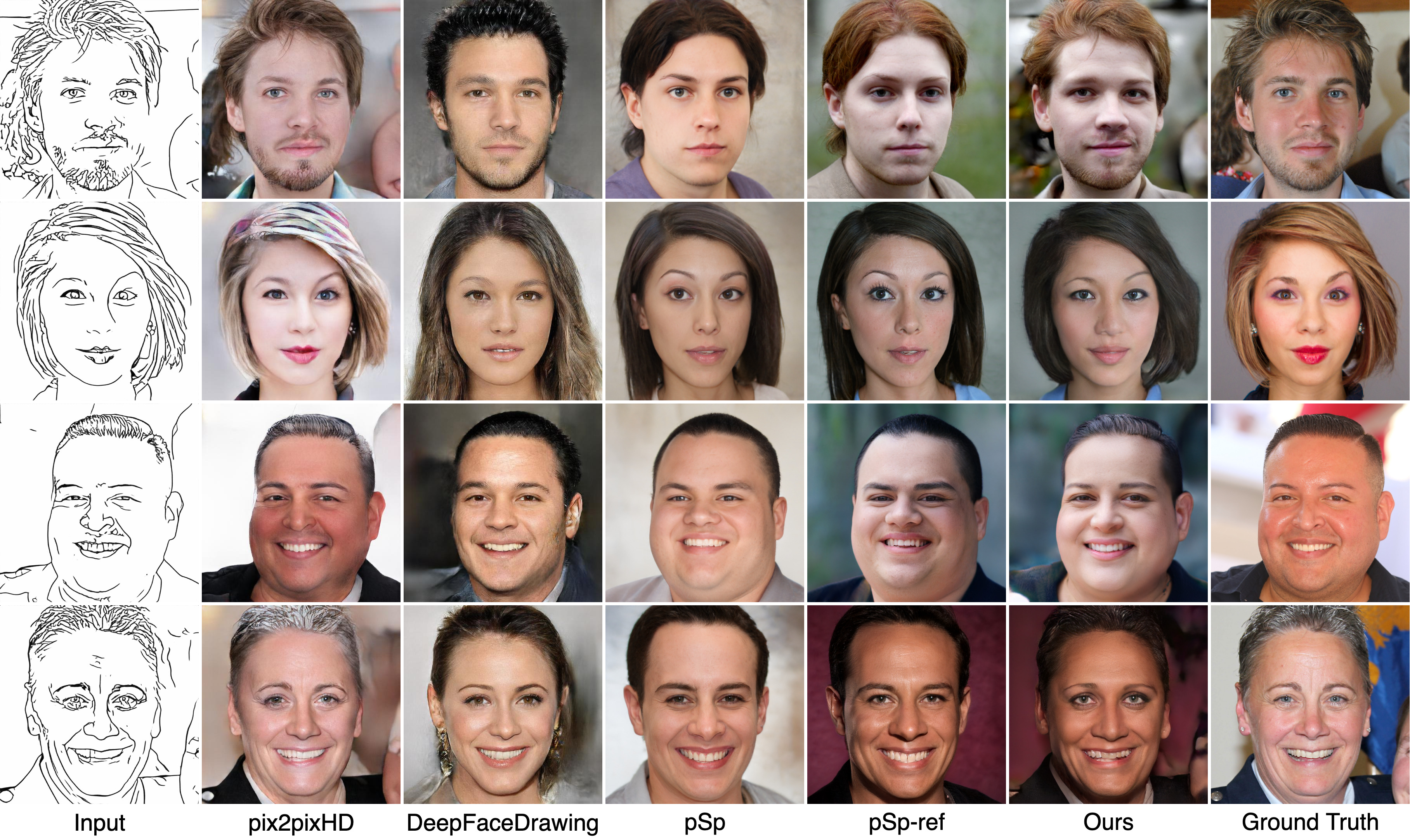} \\
    \caption{Generation comparisons with the state-of-the-art methods given the sketches extracted from the FFHQ dataset. We apply the same low-level features in both pSp-ref and ours. 
    }
    \label{fig_generation}
\end{figure*}
In Figure \ref{fig_inputs}, we can see that with only semantic map input, the synthesized results present clear region boundaries, but lack clear internal structures, see the blurry glasses and plain hair region in the two examples, Figure \ref{fig_inputs}(c).
Adding semantic map effectively eliminate the sketch ambiguity, see the cloth regions in second row, (b) and (d); With additional sketch, the boundary present sharp edges (e.g. hair boundary in second row, (c) and (d)).
Generally, the models with inputs of sketch only and sketch-map combination produce visually similar results, while additional modality of input provides additional guidance and leads to more detailed results.

To evaluate the generation ability of our \moduleName, we compared our method with several state-of-the-art image generation frameworks including pix2pixHD~\cite{wang2018high}, DeepFaceDrawing ~\cite{chenDeepFaceDrawing2020}, and pixel2style2pixel (pSp) ~\cite{richardson2020encoding}. 
We trained the mentioned methods (except DeepFaceDrawing) using their released codes, with our generated sketch-image pairs used in our \moduleName training process.
For DeepFaceDrawing, we directly input the sketches to their online system to get the resulting images, since DeepFaceDrawing is designed for generating frontal faces and re-training it on our training data (involving faces under various poses) would deteriorate its performance.
To conduct a fair comparison, we chose our architecture with the sketch input only.
The spatial resolutions in this comparison are: Ours ($512 \times 512$ input, $1024 \times 1024$ output), pix2pixHD ($512 \times 512$ for both input and output), pSp ($256 \times 256$ input, $1024 \times 2014$ output), and DeepFaceDrawing ($512 \times 512$ for both input and output).

For the qualitative comparison, we randomly selected a collection of 500 portrait images in FFHQ \cite{karras2019style} and extracted the corresponding sketches. \wanchao{Since no available style code is paired with samples in FFHQ, we adopted the randomly selected low-level styles in our style code dataset to provide the coloring and texture details in the synthesized images for qualitative evaluation. See Figure \ref{fig_generation} for the visual comparison. In this figure, the results of pix2pixHD are more similar to the ground truth in terms of the sketch correspondence, while the image quality of pix2pixHD is inferior to ours. 
In addition, due to the adoption of the StyleGAN architecture, our approach can easily change the low-level coloring scheme of the results and such effects are not feasible using pix2pixHD.}

For a quantitative evaluation, we utilized our test dataset and performed the image reconstruction task, \wanchao{since we have the ground truth images paired with style codes}. 
Here we chose pix2pixHD \cite{wang2018high} and pSp \cite{richardson2020encoding} as the comparison methods representing the state-of-the-art pixel-wise image translation and StyleGAN encoding method, respectively.
We evaluated sketches from the test set and used their paired low-level style codes (if applicable) in generation.
% We show an example for the quality comparison with the test dataset in Figure \ref{fig_overlay}.
We compared pSp with both the original generation strategy in their sketch-to-image setting with no style codes as input, as well as the style-mixed (8-18 as suggested by its authors) version of the sketch-to-image generation.
The reconstruction results were measured not only by the metrics used above, but also SSIM \cite{wang2004image} and PSNR.
The results are listed in Table \ref{tab_generation}.

\begin{table}
    \centering
    \begin{tabular}{ccccc}
    \hline
        Method & pix2pixHD & pSp & pSp-ref & Ours \\
    \hline
        L1     &  0.121 &  0.198 &  0.146 & \textbf{0.105} \\
        Local  &  0.277 &  0.397 &  0.284 & \textbf{0.190} \\
        Global &  0.105 &  0.191 &  0.135 & \textbf{0.076} \\
        FID    & \textbf{34.021} & 67.214 & 52.029 & 35.116 \\
        SSIM   &  0.448 &  0.222 &  0.353 & \textbf{0.453} \\
        PSNR   & 10.070 &  6.193 &  7.832 & \textbf{11.029} \\

    \hline
    \end{tabular}
    \caption{\label{tab_generation} Quantitative evaluation on the generation performance among different methods. pSp-ref represents the pSp generation with reference low-level style codes.}
\end{table} 

From Table \ref{tab_generation} we can see that overall our method achieves the best performance among all the methods. 
Adopting the paired low-level styles significantly improves the performance of the reconstruction task (see pSp vs. pSp-ref).
The quantitative statistics of both pSp and pSp-ref are inferior to that of pix2pixHD and ours. This may attribute to the loose correspondence between the input sketches and generated results. 
We can see that in all the evaluation metrics, pix2pixHD achieves similar performance to ours. This somewhat confirms the pixel-wise correspondence of our method. 
Although pix2pixHD performs well in the testing data reconstruction task, it fails to presents results with high quality in the qualitative evaluations, see Figure \ref{fig_generation}.
One possible reason for this phenomena may be that there exist slight difference between the training data samples and the test data extracted from FFHQ, considering pix2pixHD based methods are sensitive to inputs.

\subsection{Evaluation of the Editing Performance}\label{sec_exp_editing}
\begin{figure}
    \centering
    \includegraphics[width=0.8\linewidth]{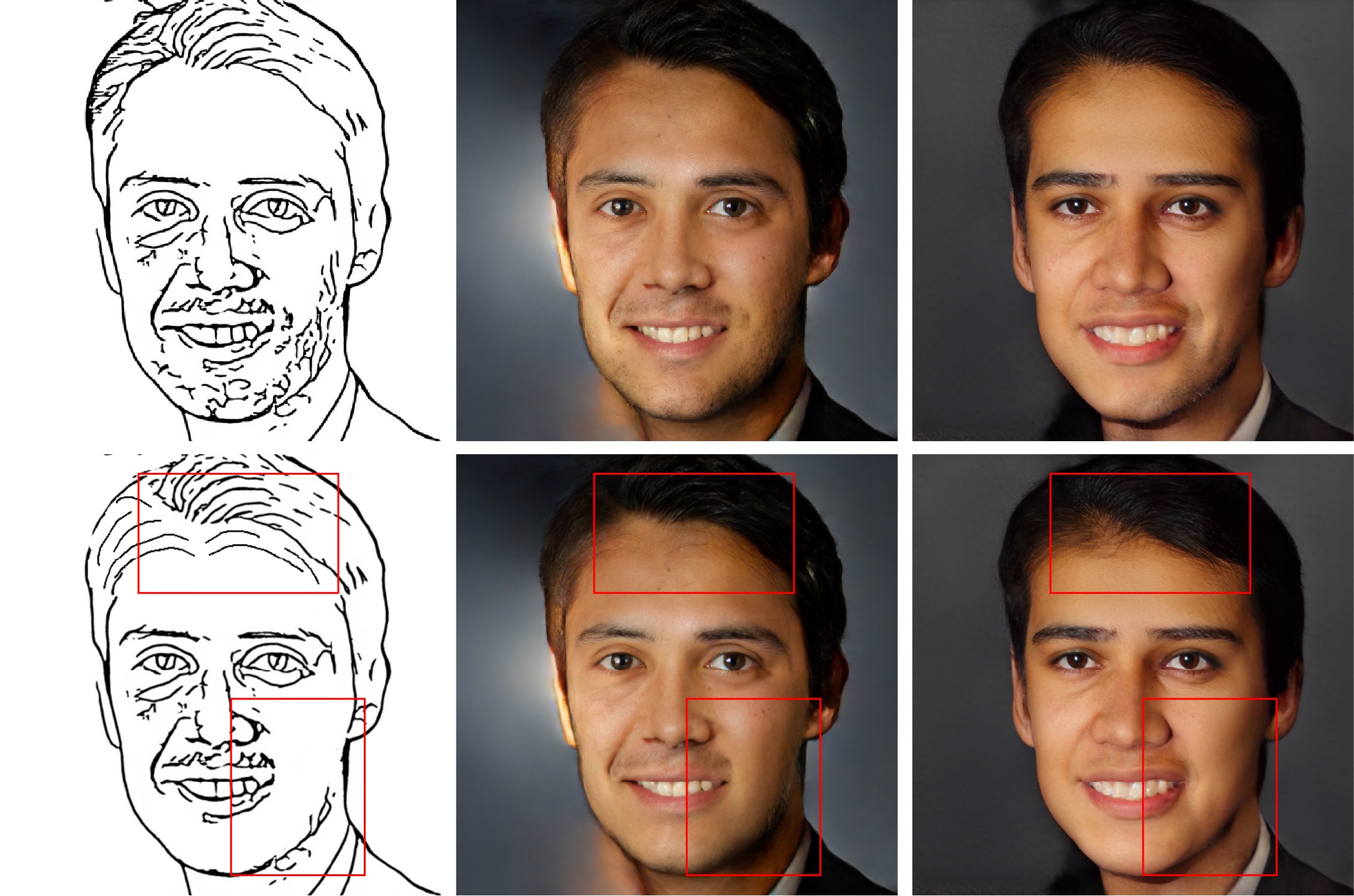}
    \caption{
    Comparisons between \emph{DeepFaceEditing} (Middle Column) and our approach (Right Column) (with a sketch only as input) on editing a front face. Top: before editing. Bottom: after editing, with the edited regions highlighted. \wanchao{Our approach leads to results with finer details and provides a better response to the edits (i.e., adding head hair and removing facial hair in this example). The difference is best viewed with zoom-in. For a fair comparison, our model is trained with only sketch inputs.}
    }
    \label{fig_editing_comp}
\end{figure}

\begin{figure}
    \centering
    \includegraphics[width=0.9\linewidth]{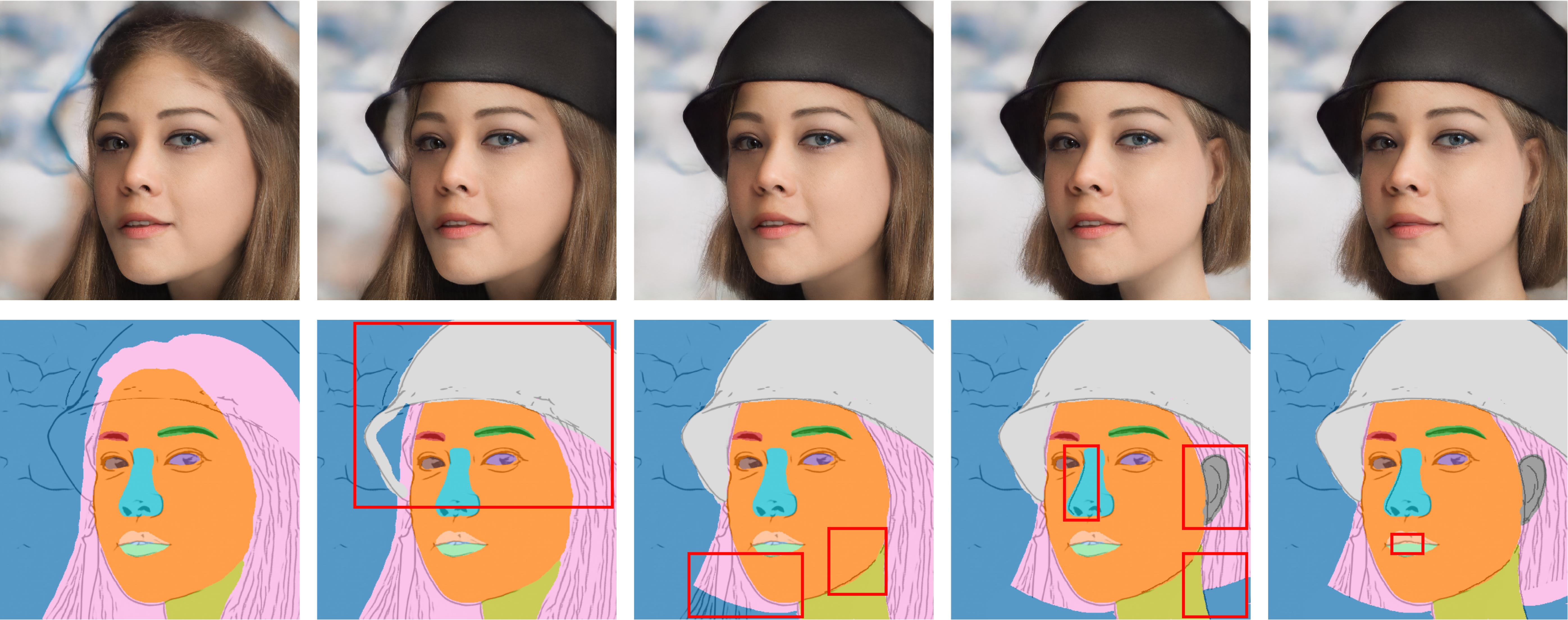}
    \caption{A series of editing operations on both semantic maps and sketches (Bottom) and corresponding synthesized results (Top). \wanchao{The edited areas are highlighted with red boxes in the bottom.}}
    \label{fig_editing_series}
\end{figure}

To show the effectiveness of the editing mode, we compared our system with a recent work \emph{DeepFaceEditing} \cite{chenDeepFaceEditing2021}, which is a state-of-the-art portrait editing technique but focuses on frontal face editing.  
Similar to \emph{DeepFaceDrawing} \cite{chenDeepFaceDrawing2020}, since \emph{DeepFaceEditing} was designed for frontal faces, we did not re-train their model but used their released code with a pre-trained model to directly test their editing results.
To conduct a fair comparison, we used the generation model trained with only sketch inputs. The comparison results are shown in Figure \ref{fig_editing_comp}.
\wanchao{We can see that due to adoption of the StyleGAN architecture, our method produces results with finer details like the skin texture and fewer artifacts (see the right ear and neck regions of the results by \emph{DeepFaceEditing}).
For adding head hair and removing facial hair, our method provides a better response than \emph{DeepFaceEditing}}.
For editing with non-frontal faces or accessories like glasses or hat, \emph{DeepFaceEditing} fails to provide high-quality results due to their component-aware design and frontal face pre-settings.

To demonstrate the full capability of our editing mode, we illustrate a series of editing operations in Figure \ref{fig_editing_series} with our full method (model trained with sketch and semantic map inputs).
We modified the image with both the sketch and semantic map to depict the desired modifications.
As mentioned in Section \ref{sec_exp_ablation}, our method can produce detailed results with only a sketch input (e.g., the eyeglasses in first case of Figure \ref{fig_inputs}(b)). 
However, such precise sketches (edge maps) are difficult for ordinary users to draw, even with the sketching assistance.
We utilize the semantic map to solve this sketch ambiguity (i.e., defining the boundary of hair and background) by directly providing the semantics of the region and leave sketch to only depict structural features, see the hat creation process in Figure \ref{fig_editing_series}.
Using both the sketch and the semantic map, drawing for generation and/or editing is greatly simplified by defining region boundary (semantic map) and adding structural details (sketch).
Adopting this principle, novice users can produce high-quality portrait images using our method with great ease.

\subsection{Usability Study}\label{sec_exp_usability}
To evaluate the effectiveness and usability of our system, we conducted a usability study, including two parts: a fix-task study and an open-ended study. We recruited 12 participants (6 female, aged from 23 to 31, U1-U12) and asked them to evaluate their drawing skills from 1 (poor) to 5 (good). 8 out of them were novice or middle users (score: 1-3). 
Each participant was requested to perform a fixed-task drawing session as the training process for using our system, followed by an open-ended drawing session to let them freely express their design ideas with our system.

\begin{figure}
    \centering
    \includegraphics[width=0.9\linewidth]{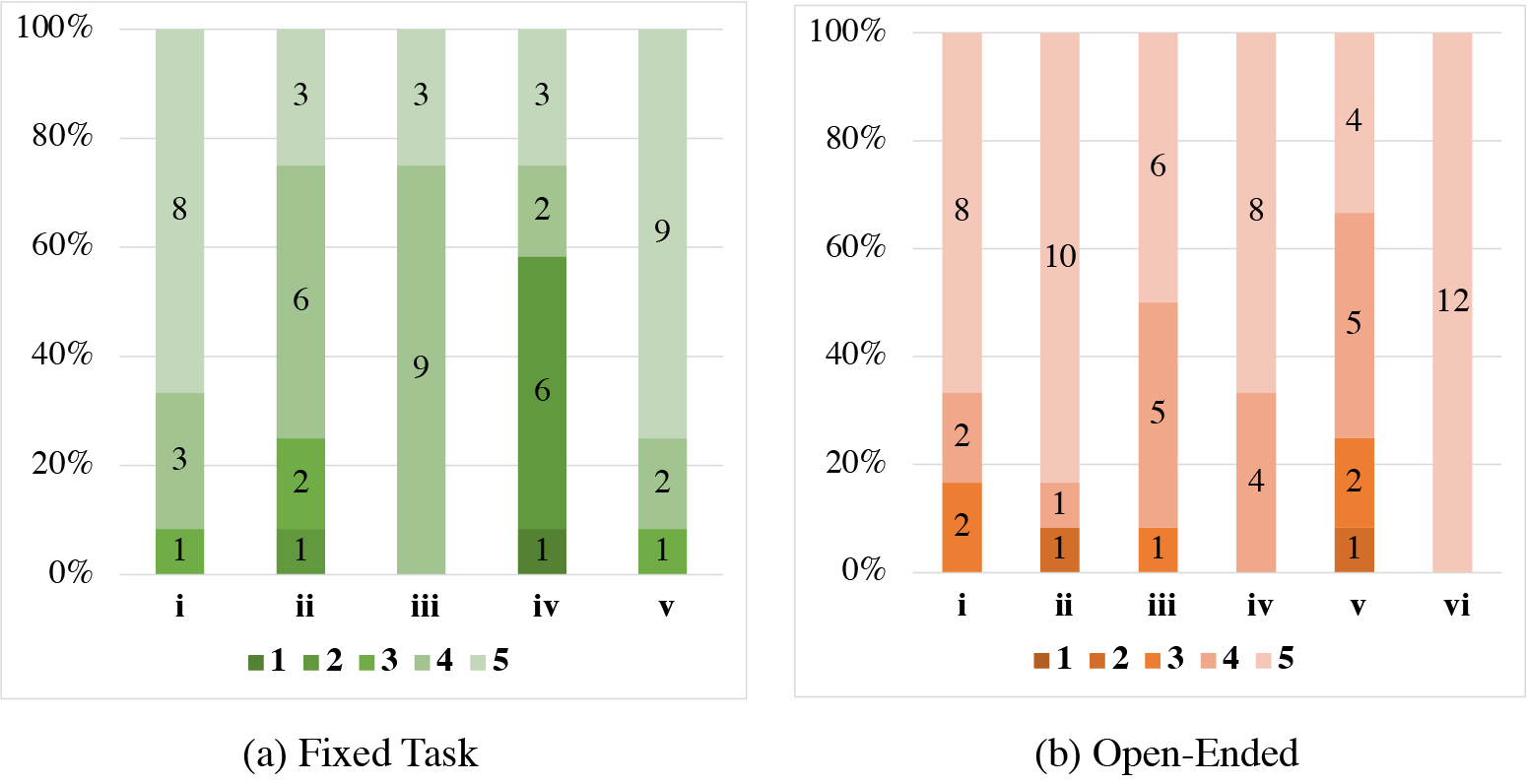} 
    \caption{The subjective ratings of the fixed-task and open-ended studies. The five colors represent the respective scores from 1 to 5. The numbers in different parts of each column are the number{s} of participants giving specific scores. i to v in (a) refer to ease of use, consistency with target portrait, precise, effort, and helpfulness, respectively. i to vi in (b) refer to result diversity, result quality, expectation fitness, helpfulness of global guidance, helpfulness of local guidance, and helpfulness of the combination of sketch and map, respectively. }

  \label{fig:sub_rating}
\end{figure}
 
\subsubsection{Fixed-task Study}
In the fixed-task study, we selected two portrait images from the test set as the target images.
% (Figure \ref{fig_fixed_task} (Left)). 
To cover different genders, poses, face components, and expressions, we selected a smiling female with frontal face and a gazing male with side face. 
We asked the participants to reproduce the portrait images using our system as similar to the target images as possible. During the study, the two target images were shown on a display in front of the participants. After they finished drawing, they were asked to fill in a questionnaire to evaluate \emph{ease of use}, \emph{consistency with target portrait}, \emph{precise control}, \emph{effort}, and \emph{helpfulness of guidance} in a 5-point Likert scale (1 = strongly disagree to 5 = strongly agree). 

Figure \ref{fig:sub_rating}(a) plots the distribution of subjective ratings on the five measures.
From the figure, we can conclude that most participants could produce their satisfied images similar to the target images easily and with precise control over the details of the face components. All of them rated the precise control as 4 or 5 point, validating the good controlibility of our system over the whole face. 5 participants rated the effort as 4 or 5, since they thought they were not very familiar with the operations and functions of our system. U6 commented that ``\emph{it took a while to learn the interface and the tools}''. Most participants (11) considered the suggestive guidance in our system very helpful in reproducing target faces.

\subsubsection{Open-ended Study}
\begin{figure*}
    
    \includegraphics[width=0.99\linewidth]{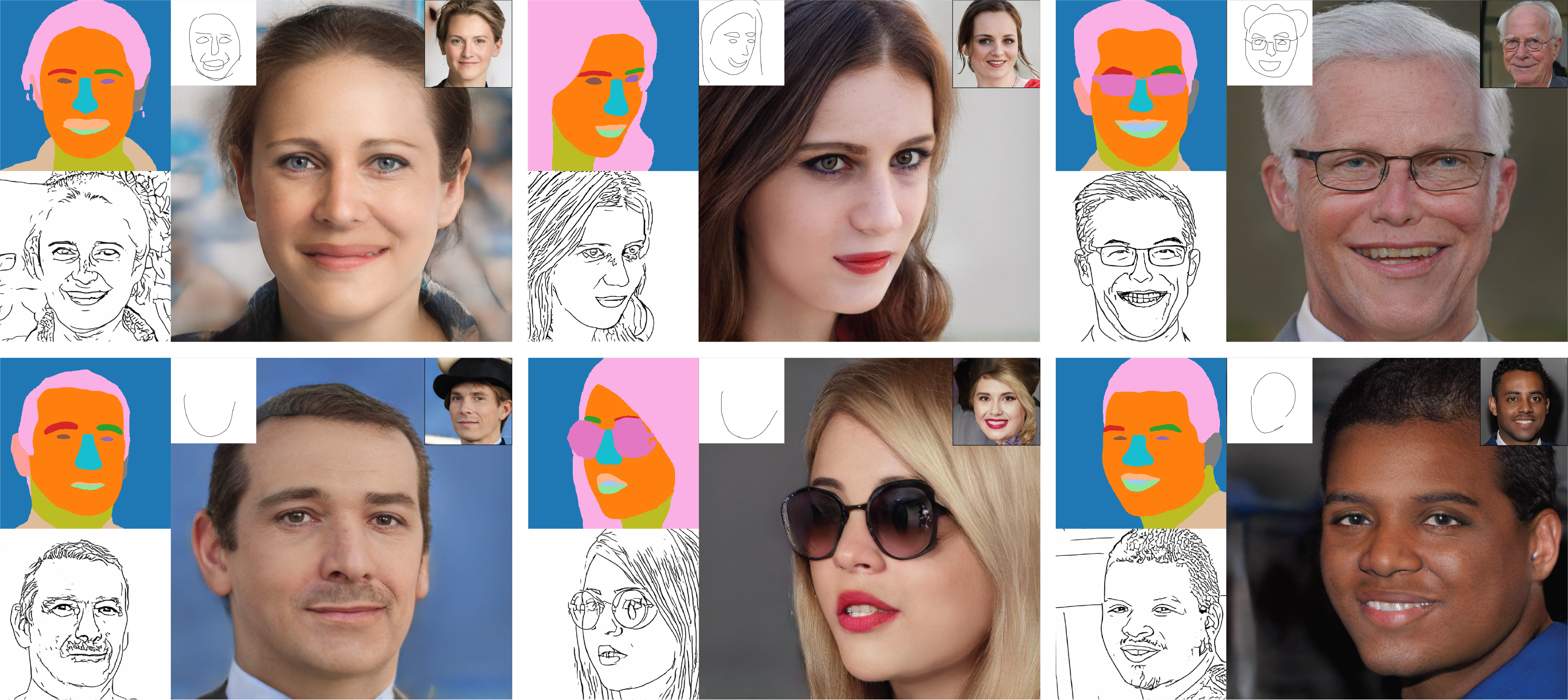} \\
    \caption{Representative results from the open-ended study. The left column of each group shows the final sketch 
    % \hbc{I'm not sure if you recorded or not, if yes, it would be great to highlight the user-drawn strokes in different colors.}
    and semantic map specified by the users, and the right image is the corresponding synthesized portrait image by our system. The corresponding initial sketches for retrieval are illustrated in thumbnails on the upper left corners of synthesized portrait images. The reference style given in thumbnail on the upper right corner of each synthesized portrait image are chosen by the users.}
    \label{fig_open_ended}
\end{figure*}

In the open-ended study, we asked the participants to create their desired face images using our sketch-based suggestive system. At the end of the study, they were asked to fill in a questionnaire to evaluate the different features of our system in a 5-point scale (1: strongly disagree to 5: strongly agree). Figure \ref{fig_open_ended} shows the representative result images by different participants with the initial user-drawn strokes as well as the refined sketches and semantic maps.
As seen in Figure \ref{fig_open_ended}, our method can
help users turn initial rough sketches into high-quality photo-realistic portrait images. Please refer to our supplemental materials for more results.

Figure \ref{fig:sub_rating}(b) plots the distribution of their ratings. From the figure, 10 out of 12 participants thought that they could produce very diversified (rating on 4 and 5) result images using our system. It resonates with the comments of the participants: U2 said ``\emph{[using this system] I can draw various faces with different characteristics and styles}''. 11 participants rated the result quality and expectation fitness as 4 or 5 point. U12 also pointed out that ``\emph{the generated face is even more beautiful than I imagined, with good quality}''. All of them found the global guidance very helpful. U5 said ``\emph{global suggestions is very useful because it helps me select a desired template according to my strokes quickly. It reduces the drawing time to a large degree}''. Local guidance is also preferred by the participants, as reflected by the high scores in Figure \ref{fig:sub_rating} (b) and users' feedback. U6 commented that ``\emph{the control over editing the sketch is really helpful in manipulating the image}''.  U12 pointed that ``\emph{I like the fact that I can change details}''. Besides these points, the participants also said ``\emph{this tool is useful for users  who have limited drawing experience}'' (U2 and U7). The participants also loved the pose selection function: U2 said ``\emph{reference in start sketch (pitch/yaw) is super useful}''. 

To further validate the different roles of sketch and map modification in our system, we asked the participants to rate on the effectiveness of sketch and semantic map modification in structure and region editing, respectively. We found that the mean scores of sketch for structure and region editing, semantic map for structure and region editing are 4.67, 4.50, 4.50, and 4.75 (SD: 0.49, 0.67, 1.00, and 0.45), respectively.
The high scores here indicated the agreement of the participants on the importance of using both sketches and semantic maps. It is also interesting to note that the score of sketch for structure editing is slightly higher than the score of sketching for region editing; the score of semantic map for region editing is slightly higher than the score of semantic map for structure editing. This indicates that the sketch is more suitable for controlling the structure (e.g., lines, curves, wrinkles, textures) while the semantic map is more useful for modifying the region (e.g., hair region, background, cloth region). These two features supplement each other {and reduce the ambiguities in the drawing}, thus facilitating the easy and controllable portrait image creation process. 

Besides the advantages of our system, the participants also provided us with some suggestions for further improvement. For example, U12 suggested a rotation function of the selected component since it is necessary to rotate a face component especially in side poses. We will incorporate both rotating and scaling functions of the face component in the future.

\subsection{Perceptive Study}\label{sec_exp_perceptive}
To compare the visual fidelity (i.e., the degree of closeness to the ground truth) of the reconstructed results with different methods (Table \ref{tab_generation}), we conducted  the first task of the perceptive study.
For the comparison with in-the-wild sources (samples from FFHQ dataset, Figure \ref{fig_generation}), since both our method and pSp provide photo-realistic results, we thus conducted the second task evaluating the generation faithfulness concerning the sketch inputs.
\begin{figure}
    \centering
    \includegraphics[width=0.8\linewidth]{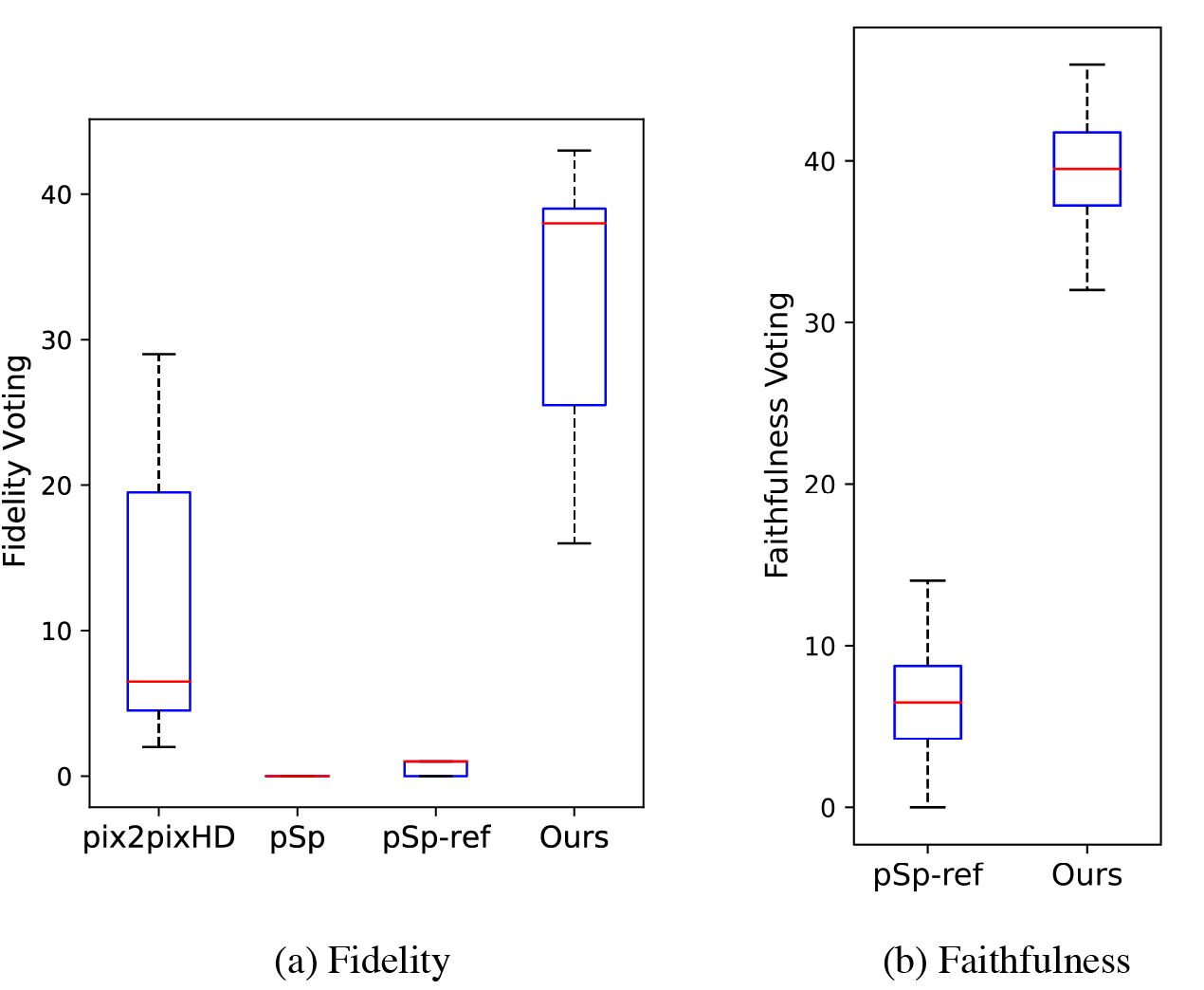}
    \caption{Box plots of the fidelity and faithfulness perception votes (averaged over the questions) over the participants for each method.}
    \label{fig_perceptive}
\end{figure}

In Task 1, we prepared a set of reconstructed results randomly selected from the test results, containing 10 samples synthesized by all the compared and the ground-truth images. 
For each trial, i.e., each group of results, we asked the participants to select the most similar candidate to the ground-truth image.
For Task 2, we provided an input sketch and two generated results by our method and pSp with the same low-level reference style.
We asked the participants to choose the candidates with the best sketch-correspondence.
We used an online questionnaire to perform this study. 46 participants (30 male, 16 female, 41 in age range 20-30) participated in this study. 
We counted the number of votes of each method in all the questions.
Figure \ref{fig_perceptive} plots the statistics of the evaluation results. 
We performed single-factor ANOVA tests on the quality and faithfulness scores, and found significant effects for both fidelity ($F_{(3,36)} = 48.74$, $p < 0.001$) and faithfulness ($F_{(1,18)} = 254.24$, $p <0.01$).

\subsection{Extension to More Categories}\label{sec_exp_other}
\begin{figure*}
    \centering
    \includegraphics[width=0.96\linewidth]{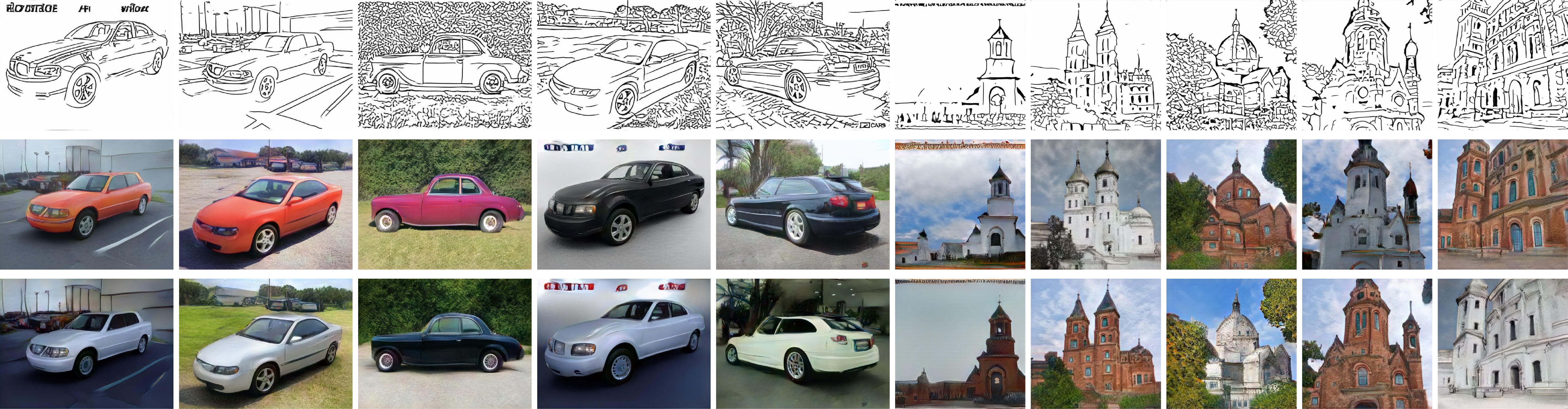}
    \caption{Representative test results. The first row illustrates the sketch inputs, and the second and third rows show the corresponding synthesized results with randomly selected reference styles. \wanchao{All the results are produced by models trained with sketch inputs only.}}
    \label{fig_car}
\end{figure*}

% \hbc{Need a short discussion about our ideas work or do not work for what kind of object categories.}
Although we focus on face image generation and editing in this work, our \moduleName~is not limited to faces. In fact, our conditioning idea can be applied to pre-trained weights on any datasets. To show this, we extend our modification to more categories of data in this subsection.
%Our main idea is to utilize the pre-trained StyleGAN and modify it as a conditional generation framework.
%\wanchao{We present the experiment, extending our modification to more categories of data to show the generalization ability of our SC-StyleGAN (i.e., our conditioning idea can be applied to pre-trained weights on any datasets).}
While we mentioned that all experiments are based on the StyleGAN2 \cite{karras2020analyzing} framework pre-trained on the FFHQ dataset, the modification is framework-irrelevant. This means our conditioning idea can be applied to pre-trained weights on any datasets, for both StyleGAN\cite{karras2019style} and StyleGAN2\cite{karras2020analyzing}.
Here we applied our proposed ideas to the architecture with weights pre-trained on LSUN \emph{Car} and \emph{Church} dataset \cite{yu2015lsun} of resolution $512\times384$ and $256\times256$, respectively.
We adopted a similar data preparation process and collected 5K data samples for each dataset (4.5K for training and 0.5K for test).
In this experiment, we used the sketch-to-image generation process and substituted the intermediate feature and image in the resolution of $32\times32$ for \emph{Car} and $16\times16$ for \emph{Church}. We illustrate the test results in Figure \ref{fig_car}.
It can be seen that although the training data are rather limited, the generated images faithfully respect the input sketches. 
With different reference styles, the results present diversified appearance while respecting the sketch guidance rigidly. 

\section{Conclusion and Discussions}

In this paper, we have presented \emph{\sysName}, a novel system to help novice users draw a photo-realistic portrait image from scratch or intuitively edit existing portrait images. 
Our data-driven suggestive interface interactively provides recommendations for global template selection and component detail refinement, and guides users to refine their drawings towards more realistic faces. To support easy depiction and precise control of generation results, we adopt two input modalities: sketches and semantic maps. 

Our novel \moduleName~takes as input a sketch and a semantic map and synthesizes a high-resolution, realistic portrait image, converting the original StyleGAN framework to an image-to-image generation architecture.
Our method outperforms the current sketch-to-portrait methods in terms of both fidelity and condition faithfulness. 
\wanchao{Using our current approach, we can strictly encode the spatial conditions in the StyleGAN generation process while preserving its generation quality, thus making our architecture possess superior StyleGAN generation capability and strict spatial condition correspondence.}
The user studies confirmed the usability and effectiveness of our system.
 
\begin{figure}
    \centering
    \includegraphics[width=0.8\linewidth]{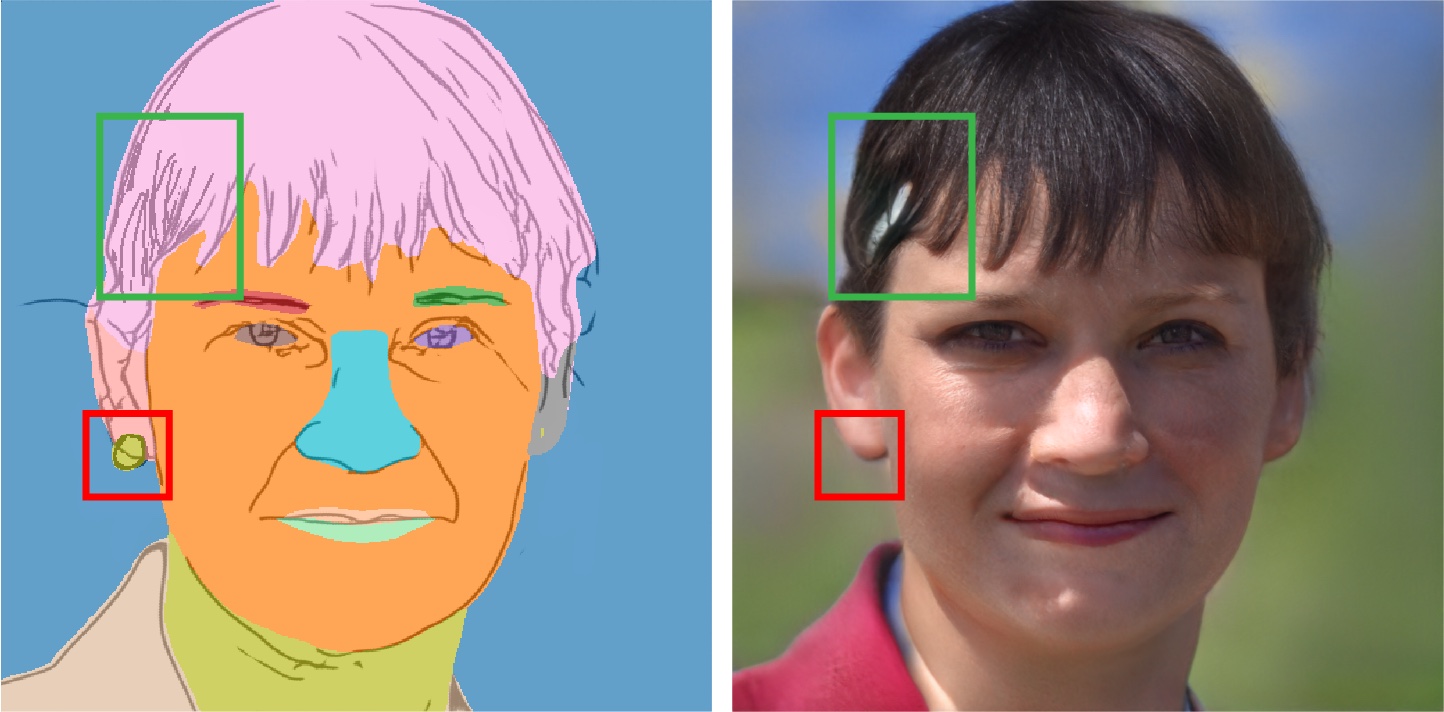}
    \caption{A less successful case. In this example, the generated result presents a visually inconsistent artifacts due to the densely sketched input. Delicate accessory like earrings can not be produced.}
    \label{fig_failure}
\end{figure}

Despite the good results produced by our system, our method might generate less successful results. Figure \ref{fig_failure} shows such an example: 
Our method responses not well to delicate accessories, like earrings and necklaces. 
Highlighted by red rectangles in Figure \ref{fig_failure}, although earring sketch and semantic map are provided, the resulting image presents no such accessory.
This is mainly because the training samples with delicate accessories are limited and often contain artifacts in the StyleGAN sampled dataset.

Another artifact shown in Figure \ref{fig_failure} is that for the densely sketched region, the generated result often synthesized it as visually unpleasing textures (highlighted in green rectangles for input and result), this is quite common in sketch-based generation methods.
In our system, we resort to the user interaction to resolve this problem, and for other non-interactive methods, providing a mechanism for replacing the densely sketched region would be beneficial.

We developed our system based on the idea of providing users with flexibility to the largest extent, this inevitably leads to cases when the semantic map and the sketch conflict with each other sometimes.
For well-aligned facial regions (e.g., eyes, eyebrows, nose, mouth, etc.), the results correspond to the semantic mask more than the sketch, since their appearances are primarily dependent on the boundary shapes. This property was often utilized by the participants when controlling the mouth/eye open/closed in the drawing session of our study.
For the other regions where the conflict often occurs, e.g., cloth, neck and background, the generated textures correspond to the sketch more, since such appearances are mainly defined by the internal textures.
\wanchao{It is also possible that the results would sometimes suffer from incompatibility of different retrieved components. 
We believe that this could be potentially resolved by a post-processing step, such as
small networks to improve the naturalness of the edited sketch and segmentation map.} 

Since our method is designed for ensuring the strict spatial correspondence between the condition and synthesized result, the non-spatial appearance extraction is not supported by our method.
In our editing session, we resorted to the existing StyleGAN inversion method like pSp to obtain the low-level appearance for the in-the-wild image editing.
Otherwise, we can only conduct editing on the images paired with style codes (e.g., the test and training samples), if the user requires to preserve the original appearance.

Our novel \moduleName~encodes the spatial condition directly to the pre-trained StyleGAN synthesis procedure instead of the widely adopted inverting-to-style code approaches.
We provide a new idea of transforming the pre-trained StyleGAN into a conditional setting, which also benefits efficient spatial embedding in the StyleGAN-based applications. 
Compared to the compact style code, our encoded feature map preserves more spatial information, thus providing results with higher spatial faithfulness to the inputs. 
Image-to-image translation applications (e.g., face super-resolution, face inpainting) attempting to utilize the pre-trained StyleGAN synthesis module could benefit from our idea. 
The reference low-level styles could be obtained from a separate branch extending from the spatial encoding module, similar to pSp~\cite{richardson2020encoding}.

One possible direction worth exploring is generating new sketches and layouts for the sketch refinement procedure, instead of directly retrieving such examples from databases. This might provide a richer set of suggestions with more details.
Sketches alone mainly provide the shape information of target faces. Currently we use reference images to control the appearance of synthesized results. In the future, it might be interesting to explore other more direct approaches to control the local appearance (e.g, via user-specified color strokes ~\cite{sangkloy2017scribbler}). 

\section*{Acknowledgement}
%\begin{acks}
We thank the anonymous reviewers for the constructive comments and the user study participants for their help. 
The work was supported by unrestricted gifts from Adobe and grants from the Research Grants Council of the Hong Kong Special Administrative Region, China (Project No. CityU 11212119), and the Centre for Applied Computing and Interactive Media (ACIM) of School of Creative Media, City University of Hong Kong.
Shu-Yu Chen was supported by the National Natural Science Foundation of China (No. 62102403). Lin Gao was supported by Beijing Municipal Natural Science Foundation for Distinguished Young Scholars (No. JQ21013), Youth Innovation Promotion Association, Chinese Academy of Sciences.
\ifCLASSOPTIONcaptionsoff
  \newpage
\fi



\bibliographystyle{IEEEtran}
% \bibliography{IEEEabrv,}
\bibliography{ref_file}
\end{document}

\title{Supplemental Materials\\\sysName: Portrait Image Generation and Editing with Spatially Conditioned StyleGAN}
\maketitle

\section{Data Preparation}\label{sec_data}

% To train our model as a conditional generation framework, we need a large-scale dataset of condition-portrait pairs.
% % \hbc{refer to some illustration figure?}. 
% A sketch and a semantic map together form the conditions to guide portrait image generation.  
% We take advantage of the generation ability of the StyleGAN framework (StyleGAN2 \cite{karras2020analyzing} throughout this paper) in constructing the training dataset by collecting a large series of generation results.
% We first sample a large collection of random vectors from a normal distribution before feeding them to the mapping network {of StyleGAN}. 
% We then input the resulting latent style codes from the mapping network to the synthesis network and obtain the portrait images corresponding to the style codes.
% Up to now, we get a collection of pairs of latent codes and images.
We resort to the sketch extraction method in \emph{DeepFaceDrawing}~\cite{chenDeepFaceDrawing2020}, which uses the PhotoCopy filter in PhotoShop and a sketch simplification method~\cite{SimoSerraSIGGRAPH2016}, to process the generated portrait images and thus get the sketch-image pairs.
For the construction of the semantic map repository of the generated portrait images, we utilize an image segmentation network, BiSeNet \cite{yu2018bisenet}, pre-trained on the CelebAMask-HQ dataset \cite{CelebAMask-HQ}.
Finally, we get 45K (40K for training) data (image-sketch-semantic map-latent code) samples for training our \moduleName~ module. 

To assist users of our system in sketching, we provide both the global face templates and the local component details in the suggestive sketching interface.
To provide the global sketching guidance, we use a subset (17K samples) of the FFHQ dataset \cite{karras2019style} and extract the corresponding sketches as \emph{DeepFaceDrawing}~\cite{chenDeepFaceDrawing2020} to construct the sketch template repository.
We also predict the semantic maps of the 17K images using the same scheme mentioned above as the semantic map dataset and extract the contours of the semantic maps to provide a retrieval intermediate in the initial global sketching stage.

For the local detail suggestion, we choose to use the CelebAMask-HQ dataset ~\cite{CelebAMask-HQ} as the component sketch candidates, since CelebAMask-HQ contains a relatively larger volume (30K) of accurate semantic maps.
We roughly divide each semantic mask into eight regions of interest, namely, ``left eye'', ``right eye'', ``nose'', ``mouth'', ``facial skin'', ``glass'', ``hat'', and ``hairs'' to embed the individual component region details (Figure \ref{fig_sketch_decompose}).  
We resort to the same sketch extraction method as mentioned above to get the component sketches for constructing the component sketch repositories for the local refinement.
To reduce the overlap between different semantic regions, we further extract each sketch image with guidance provided by a dilated region mask instead of cropping with a rectangle region \cite{chenDeepFaceDrawing2020}, and record the corresponding sketch and semantic mask.
In this way, we get clean component sketches, which reduce the conflict when replacing with the original component sketch. 
In addition, we record the portrait pose attribute ($[yaw, pitch, roll]$ triplets) of each image in both the FFHQ subset and CelebAMask-HQ dataset for pose selection.

\begin{figure}[htb]
    \centering
    \includegraphics[width=0.9\linewidth]{fig/sample-new.pdf}
    \caption{An illustration of our local detail sketch extraction process. After getting a sketch from a real image, we decompose the sketch with respect to the dilated semantic masks of inner components (``left eye'', ``right eye'', ``nose'' and ``mouth'' and ``glass'') to reduce the conflict among them when assembling.
    We extract the other component regions according to their original masks.
    } 
    \label{fig_sketch_decompose}
\end{figure}

\begin{figure}
    \centering
    \includegraphics[width=0.3\linewidth]{fig/comp-dfd/10370.png}
    \includegraphics[width=0.3\linewidth]{fig/comp-dfd/save-10370.jpg}
    \includegraphics[width=0.3\linewidth]{fig/comp-dfd/10370_12473.png}
    \caption{\wanchao{Visual comparison on the side face generation between DeepFaceDrawing \protect\cite{chenDeepFaceDrawing2020} (Middle) and our method (Right) given the same sketch input (Left). Here our method is trained with sketch only.}}
    \label{fig:my_label}
\end{figure}

\begin{figure}[t]
    \setlength{\fboxrule}{0.1pt}
    \setlength{\fboxsep}{-0.01cm}
    \begin{tabular}{ccc}
        \includegraphics[width=0.3\linewidth]{fig/open_ended/11-1.png} &
        \includegraphics[width=0.3\linewidth]{fig/open_ended/11-2} &
        \includegraphics[width=0.3\linewidth]{fig/open_ended/11-3}\\
        \includegraphics[width=0.3\linewidth]{fig/open_ended/12-1.png} &
        \includegraphics[width=0.3\linewidth]{fig/open_ended/12-2} &
        \includegraphics[width=0.3\linewidth]{fig/open_ended/12-3}\\
        \includegraphics[width=0.3\linewidth]{fig/open_ended/7-1.png} &
        \includegraphics[width=0.3\linewidth]{fig/open_ended/7-2} &
        \includegraphics[width=0.3\linewidth]{fig/open_ended/6-3}\\
        \includegraphics[width=0.3\linewidth]{fig/open_ended/5-1.png} &
        \includegraphics[width=0.3\linewidth]{fig/open_ended/5-2} &
        \includegraphics[width=0.3\linewidth]{fig/open_ended/5-3}\\ 
        \includegraphics[width=0.3\linewidth]{fig/open_ended/13-1.png} &
        \includegraphics[width=0.3\linewidth]{fig/open_ended/13-2} &
        \includegraphics[width=0.3\linewidth]{fig/open_ended/13-3}\\ 

 \end{tabular}
    \caption{Other results from the open-ended study. The left, middle, and right column of each group shows the final sketch, semantic map specified by the users, and the corresponding synthesized portrait image by our system. A reference style image is given in thumbnail on the upper right corner of each synthesized portrait image.}
    \label{sup_open}
\end{figure}
\bibliographystyle{IEEEtran}
\bibliography{ref_file}